\pdfoutput=1 
\documentclass[acmsmall,nonacm]{acmart}

\AtBeginDocument{%
  \providecommand\BibTeX{{%
    \normalfont B\kern-0.5em{\scshape i\kern-0.25em b}\kern-0.8em\TeX}}}

\renewcommand\footnotetextcopyrightpermission[1]{} 
\usepackage{booktabs}
\usepackage{xcolor}
\usepackage{tabu}
\usepackage{arydshln}
\usepackage{multicol, multirow}
\usepackage{graphicx}

\begin{document}

\title{Understanding Teenage Perceptions and Configurations of Privacy on Instagram}

\author{Dora Zhao}
\email{dorothyz@alumni.princeton.edu}
\author{Mikako Inaba}
\email{minaba@princeton.edu}
\author{Andrés Monroy-Hernández}
\email{andresmh@cs.princeton.edu}
\affiliation{%
  \institution{Princeton University}
  \city{Princeton}
  \state{New Jersey}
  \country{USA}
}

\renewcommand{\shortauthors}{Zhao et al.}
\renewcommand{\shorttitle}{Understanding Teenage Perceptions and Configurations of Privacy on Instagram}
\begin{abstract}
As teenage use of social media platform continues to proliferate, so do concerns about teenage privacy and safety online. Prior work has established that privacy on networked publics, such as social media, is complex, requiring users to navigate not only the technical affordances on the platform but also interpersonal relationships and social norms. We investigate how teenagers think about privacy on the popular image-sharing platform, Instagram. We draw on an online survey (N=144) and semi-structured interviews (N=21) with teenagers, ages 13-19, to gain a better understanding how teenagers configure privacy on the popular image-sharing platform Instagram and why they make these privacy decisions. Finally, based on our findings, we provide design recommendations towards the design of better privacy controls for promoting teenage safety online. 
\end{abstract}



\maketitle
\section{Introduction}
Social media platforms are an increasingly crucial part of the day-to-day lives for many teenagers. As of 2018, $97\%$ teenagers in the United States are on these platforms, and $45\%$ of teenagers report that they are online on a near-constant basis~\cite{pew2018teens}. As more parts of teenagers' lives are conducted on social media, understanding privacy on these platforms becomes even more important. For teenagers today who have grown up with access to digital technology, they are accustomed to navigating the complex landscapes of social media platforms, on which privacy is not solely determined by individuals and their actions but rather collectively managed by content posters and viewers~\cite{marwick2014networked, boyd2010social, petronio2002boundaries}. 

To achieve privacy, social media users must navigate both built-in features offered by platforms as well as self-configured privacy practices, such as creating secondary accounts. While there is a rich body of literature exploring how and why users manage their privacy on social media, most of these studies~\cite{acquisti2006imagined, trieu2020private, dewar2019finsta, taber2020finsta, xiao2020random, hargittai2016can, duffy2019you, salter2016privates, tufekci2012facebook} have focused on young adults (18+) as opposed to teenagers, which we define as individuals between the ages of 13-19. Of the work that has explored teenagers' privacy on social media, key factors that emerged include the ability to limit access to content and to utilize strategic vagueness, as well as the importance of social norms in influencing privacy behaviors. However, these studies have explored privacy either broadly on social media, not looking into any platform in particular~\cite{marwick2014networked, xie2015see, agosto2017don, adorjan2019new, de2020contextualizing}, or on Facebook~\cite{feng2014teens, hofstra2016understanding, dhir2016adolescents, chou2021teens}-- a platform that is no longer the most commonly used by teenagers~\cite{pew2018teens}.

We seek to understand teenagers' privacy perceptions and management techniques on visual content (i.e., images, videos) sharing platforms. Visual platforms are increasingly popular, especially with younger users. In this work, we focus on privacy perceptions on Instagram-- a popular visual social media platform for teenagers. A 2018 Pew report~\cite{pew2018teens} found that $72\%$ of teenagers, ages 13 to 17, are on Instagram as opposed to the $51\%$ that use Facebook. 

Through a series of semi-structured interviews with teenagers (ages 13-19), we investigate how and why teens configure privacy on Instagram by addressing the following questions: 

\quad \textbf{RQ1:} How do teenagers interpret privacy on visual content platforms and who are they interested in getting privacy from?

\quad \textbf{RQ2:} How do teenagers navigate privacy on Instagram? Do they use built-in privacy settings or other technical features?

\quad \textbf{RQ3:} Why and how do teenagers make these decisions about privacy?

In this paper, we contribute to understandings of teenage privacy practices in two ways. To start, we corroborate prior research findings, following a movement within the community emphasizing the importance of replication. Consistent with Marwick and boyd's work on networked privacy, we found avoiding in-person drama with peers and direct supervision, typically from adults~\cite{marwick2014networked}, motivates teenagers' desires for privacy. Further, similar to prior work~\cite{marwick2014networked, marwick2011tweet, dennen2017context}, we confirmed teenagers are also adept at using privacy features to avoid context collapse, which occurs when multiple audiences are combined into one. As a result, users face difficulty navigating between versions of themselves and maintaining authenticity across their different audiences~\cite{marwick2011tweet}. We also confirmed previous findings on teenagers' use of strategic vagueness, e.g., ``vague posting''~\cite{marwick2014networked, oolo2013performing}. 

In our work, we also provide novel insights on visual content-sharing platforms, a relatively underexplored area. We discovered that teens' understandings of trust have become more expansive, compared to prior works, which is in part attributable to the asymmetric nature of Instagram's follower-following relationships. Further, we added to the concept of ``networked privacy'' by suggesting that privacy is configured not only within but also across the different platforms in a user's social media ecosystem. Finally, we found that, while social norms can often lead teenagers to engage in more privacy risky behavior, peer-to-peer pathways are important channel for spreading awareness about different privacy features. From our findings, we posit that the current teenage safety features from Instagram and social media platforms are insufficient as they center adult supervision over teenager agency. Instead, we recommend that social media platform designers take into account how essential peer influence is for adopting privacy practices and create safety features that give users more flexibility.  

\section{Related Work}
In this section, we describe work related to privacy management, paying particular attention to teenagers' perceptions and privacy on visual content platforms.

\footnotesize
\begin{table}[]
    \centering
    \begin{tabular}{|c|p{2.75cm}p{1cm}p{3cm}p{6cm}|}
        \hline
         &Author (Year) &  Platform(s) & Study sample & \leavevmode\color{black}Key findings \\ \hline
         \multirow{3}{*}[-0.75ex]{\rotatebox[origin=c]{90}{\textbf{Perceptions of privacy}}} & 
         Marwick and boyd (2014)~\cite{marwick2014networked} & SNS & Semi-structured interviews with 166 US teenagers (ages 13-19) & \leavevmode\color{black}Proposed a new way for understanding privacy on social media: networked privacy\\
         
         &Agosto and Abbas (2017)~\cite{agosto2017don} & SNS & Survey and focus group with 98 US high school seniors (ages 18-19) & \leavevmode\color{black}Learned teens feel tension between having unintended audiences see their content but still want to share information with their friends \\   
         
         &Adorjan and Ricciardelli (2019)~\cite{adorjan2019new} & SNS & 35 focus groups with 115 Canadian teenagers (ages 13-19) & \leavevmode\color{black}Found teenagers claimed to ``have nothing to hide'' online while still implementing sophisticated privacy management techniques\\ 
         
         \hdashline
         \multirow{9}{*}[-19.5ex]{\rotatebox[origin=c]{90}{\textbf{Predictors of privacy}}} & Livingstone (2008)~\cite{livingstone2008taking} & MySpace, \newline Facebook, Bebo, Piczo & Open-ended interviews with 16 UK teenagers (ages 13-16) & \leavevmode\color{black}Found that older teens' presence online was less decorative and more focused on authentic relationships compared to younger teens\\
         
         &Feng and Xie (2014)~\cite{feng2014teens} & Facebook & Phone interviews with 802 US teenage Facebook users (ages 12-17) & \leavevmode\color{black}Found that parents' level of privacy concern, SNS usage, and parents' education level are positively correlated with teens' privacy concerns\\
        
         &Xie and Kang (2015)~\cite{xie2015see} & SNS & Survey of 800 US teenagers (ages 12-17) and their parents & \leavevmode\color{black}Discovered that demographics (e.g., gender, age) and SNS usage influence the amount of personal information teens share on social media \\   

         &Dhir et al. (2016)~\cite{dhir2016adolescents} & Facebook & Survey of 380 Indian Facebook users (ages 12-18) & \leavevmode\color{black}Found that gender, extroversion, and perceptions about the publicity of online photos are correlated with likelihood of untagging photos\\ 
         
         &Hofstra et al. (2016)~\cite{hofstra2016understanding} & Facebook & Large-scale survey of 3,434 Dutch high school students over three years starting at ages 14-15 & \leavevmode\color{black}Learned that peer influences, popularity, gender, and ethnicity are predictors of teenager's privacy settings\\    
         
         &De Wolf (2020)~\cite{de2020contextualizing} & SNS & Survey of 2000 Belgian students (ages 11-21) & \leavevmode\color{black}Proposed network defeatism, the ``feeling of fatalism'' about individual control on SNS influences privacy decisions\\
         
         &Kang et al. (2021)~\cite{kang2021teens} & Douyin & Survey of 500 Chinese teenage Douyin users & \leavevmode\color{black}Discovered discussion-based (compared to rule-based) parental mediation leads to more careful decisions about privacy from teens\\    
         
         &Chou and Chou (2021)~\cite{chou2021teens} & Facebook & Survey of 1956 Taiwanese students (ages 13-19) and analysis of their Facebook accounts & \leavevmode\color{black}Found that perceived vulnerability, gender, and perception of self-efficacy mediate whether teens proactively or reactively engage in privacy-protecting behavior\\         
         \hdashline
         \multirow{4}{*}[-1.5ex]{\rotatebox[origin=c]{90}{\textbf{Privacy management techniques}}}
         & Oolo and Siibak (2013)~\cite{oolo2013performing} & Facebook,\newline Twitter, blogs, IM & Semi-structured interviews with 15 Estonian Internet users (ages 13-16) & \leavevmode\color{black}Found teens use social steganography, self-censorship, and strategic information sharing to manage privacy online\\
         
         &Balleys and Coll (2017)~\cite{balleys2017being} & Facebook, Ask.fm & Online ethnographic study on accounts of teenage users (ages 14-17) & \leavevmode\color{black}Discovered teenagers use public displays of intimacy as a means of producing social capital\\
         
         &Dennen et al. (2017)~\cite{dennen2017context} & SNS & Three-day observations of 10th and 12th grade classrooms in the US, surveys, and worksheet activities & \leavevmode\color{black}Found teenagers are aware of and proactively use methods to avoid context collapse\\
         
         &Yau et al. (2019)~\cite{yau2019s} & Facebook,\newline Instagram & Ten focus groups with 51 US students (ages 12-18) in Southern California & \leavevmode\color{black}Learned teens invest effort, including asking friends for help, to select and share content to appear more interesting as to gain peer approval\\
         \hline
    \end{tabular}
    \caption{Previous literature on teenager privacy on social media. We define ``teenager'' as being between the ages of 13 and 19. ``SNS'' indicates studies that study privacy on social networking sites more broadly i.e., privacy on ``social media'' generally as opposed to specific platforms.}
    \label{tab:lit_review}
\end{table}
\normalsize
\subsection{Managing Online Privacy on Social Media} 
We ground our study in Petronio's foundational theory of Collective Privacy Management (CPM)~\cite{petronio2002boundaries}: online information is co-owned and privacy boundaries collectively coordinated. Especially on social media platforms, trying to exercise individual control over online content is particularly difficult for users~\cite{hargittai2016can}. This is because platforms are defined by the following affordances: persistence, replicability, scalability, searchability, and shareability~\cite{boyd2011social, papacharissi2011fifteen}. Thus, a prevailing model for understanding privacy on these networked platforms is through the framework of ``networked privacy''~\cite{marwick2014networked, palen2003unpacking, boyd2014s, salter2016privates, cho2016networked}. Proposed by Marwick and boyd~\cite{marwick2014networked}, networked privacy suggests that individuals no longer have sole control over their own privacy on these networked platforms. Rather, privacy is affected as much by the context in which individuals operate and social norms as it is by the technical affordances offered on the platform. Managing privacy on social media is a collective, not an individualistic, process. 
Our work aims to add to this literature in two-fold. First, while there are many studies looking at the privacy management of adults (ages 18 and older) on social media~\cite{stutzman2010friends, yang2016exploring, cho2018collective, mansour2021collective, hargittai2016can, tufekci2008can}, there has been less of a focus on teenagers (ages 13-19). This age group is particularly at risk for privacy violations on social media platforms such as being sent explicit content~\cite{pew2018, wells2021fb, mchugh2018social}. Second, for works that do focus on younger users~\cite{marwick2014networked, boyd2014s}, we aim to update our understanding of teenagers' privacy perceptions and take into account changes in social media practices, such as the shift to more visual content-sharing platforms~\cite{pew2018teens}. 

\subsection{Teenage Perceptions of Online Privacy}
Early understandings of teenage privacy centered around the “privacy paradox,” referring to the disconnect between teenagers' concerns about privacy and their lack of privacy-protective behavior~\cite{barnes2006privacy}. However, recent studies have complicated this understanding of privacy, showing that teenagers do engage in privacy-protecting behaviors~\cite{marwick2014networked, dhir2016adolescents, adorjan2019new, yau2019s}. As shown in Tab.~\ref{tab:lit_review}, we find that prior work on teenager privacy on social media falls under three broad categories: predictors of privacy, perceptions of privacy, and privacy management techniques. 

First, there is body of research focusing on understanding what factors influence teenagers' differing opinions of privacy online. In part, demographic factors influence the extent to which teenagers manage their privacy. Age and gender are commonly found to correlate with privacy management~\cite{hofstra2016understanding, livingstone2008taking}. Specifically, girls~\cite{de2020contextualizing, hofstra2016understanding, feng2014teens, chou2021teens, agosto2017don} have been found to apply more proactive and stricter privacy management practices. {In contrast, Dhir et al.~\cite{dhir2016adolescents} found that males were more likely to untag photos. Dhir et al. explain that this may because males are more likely to disclose personal information~\cite{xie2015see} and less likely to anticipate potential privacy threats~\cite{fogel2009internet}, resulting in the need to untag themselves from more posts.} Other factors that mediate privacy management practices include parental influence~\cite{kang2021privacy, feng2014teens}, personality traits (e.g., extroversion)~\cite{dhir2016adolescents}, and the nature of content being shared~\cite{de2020contextualizing}. Our work focuses less on \textit{what} factors may influence teenagers' understandings of privacy and more so on how they are understanding and configuring privacy. 

Our work is more closely aligned with prior studies that have attempted to understand how teenagers perceive privacy~\cite{marwick2014networked, agosto2017don, adorjan2019new}. We build upon the concept of ``networked privacy'' from Marwick and boyd~\cite{marwick2014networked}, who proposed that teenagers understand they must navigate changing contexts and interpersonal relationships to maintain their privacy on networked platforms, such as social media. Within this complex landscape of networked privacy, teenagers must grapple with wanting to share personal information with their friends and establishing an identity on social media while preventing unwanted audiences from seeing their content~\cite{agosto2017don, livingstone2008taking}. To navigate these networked publics, teenagers make use of a variety of privacy techniques, including vagueposting~\cite{oolo2013performing}, direct messaging~\cite{balleys2017being}, and secondary accounts~\cite{dennen2017context}. In our work, we update the understandings of teenager perceptions of privacy on social media to encompass visual content platforms. While there have been studies on visual platforms~\cite{xiao2020random, taber2020finsta, dewar2019finsta, choi2018instagram, rashidi2020s, trieu2020private, salter2016privates, duffy2019you}, this work focuses on an older population (ages 18 and older). In fact, teenager privacy on visual content platforms, which are increasingly popular with these younger users~\cite{pew2018teens}, is largely underexplored.

\subsection{Privacy on Visual Platforms}
Social media platforms offer built-in privacy features. Most of these features are designed for helping the user manage the audience of their content~\cite{duffy2019you, wisniewski2016framing}. For example, on Instagram, users are able to change account visibility from public to private, control account followers, and determine who is allowed to comment or see certain posts. Prior work has also explored how built-in features, such as ``untagging''~\cite{dhir2016adolescents, lang2015just, strano2013covering}--removing a tag on a shared photo from another user-- and ephemeral content~\cite{trieu2020private} (i.e., Instagram Stories), are utilized as tools for achieving privacy. 

In addition to making use of default features available on social media, users will reconfigure parts of the platform or make use of alternative privacy practices that fall outside these default options. One common line of inquiry within the literature is exploring ``finstas,'' which are private, secondary Instagram accounts. Prior works~\cite{xiao2020random, taber2020finsta, dewar2019finsta} have found that finstas are primarily used to post content that is not considered acceptable on primary accounts. As opposed to primary Instagram accounts, which are heavily curated~\cite{xiao2020random, yau2019s}, users are more likely to share unedited images, risqu\'{e} content, or difficult emotions on their finstas. Typically, these self-configured spaces are available to a more select audience, allowing users to create more authentic spaces online. 

Outside of studies on finsta, little work has been done to understand other ways users self-configure privacy on social media. In addition, prior studies have analyzed built-in features~\cite{dhir2016adolescents, lang2015just, strano2013covering, trieu2020private} and self-configured practices in isolation. However, in practice, users use the different tools at their disposal in tandem to achieve privacy. In this work, we provide a more holistic exploration into how teenage users manage their privacy. Specifically, we consider how teenagers use built-in features-- on and across different social media platforms, self-configured practices, and offline interactions.

\section{Methods}
\label{sec:methods}
Our study included both an online survey (N=144) and semi-structured interviews (N=21). To be eligible for our study, participants needed to be between the ages of 13 and 19, the standard age range for teenagers, and have had an Instagram account for more than 6 months, as we wanted participants who were familiar with the platform. Our university human subjects ethics board approved this study.

\subsection{Procedure}
First, participants were directed to a survey that included general questions about their usage pattern (e.g., number of accounts, how often they used Instagram), details about what features they used, and participant demographics. The survey's purpose was two-fold: to filter out individuals who did not meet our participant criteria (described above) and to better gauge popular privacy features as to guide our semi-structured interviews (see Appendix~\ref{sec:interview} and~\ref{sec:survey}). In addition, at the end of the survey, we asked whether participants wanted to be interviewed. For our first ten interviews, we reached out to all the survey respondents, and for subsequent interviews, we selectively recruited interview participants to collect a demographically representative sample.   

To recruit our participants, we used a variety of methods: posting on online forums for teenagers (e.g., Subreddit r/trueteenagers), posting flyers in high traffic areas for teenagers (e.g., cafes, restaurants), reaching out to high school administrators or teachers, and contacting teenagers the authors knew. As to not only attract privacy-conscious participants, we described our study as being focused on ``teen social media usage.'' Of the 144 individuals who completed our survey, 54 expressed interest in being interviewed, and we were able to conduct semi-structured interviews with 21. 

We conducted 45-60 minute interviews virtually from October 2021 to April 2022. We compensated participants with a $\$15$ gift card. The aim of our interviews was to understand how teenagers perceive and maintain their privacy on Instagram as well as what privacy threats they faced. Participants were asked questions about how they limited their audience and the content that they post on Instagram. This often led to discussion of specific features, both built-in (e.g., ephemeral stories, Close Friends list) and self-configured (e.g., multiple accounts), that participants use to achieve privacy. While our interviews centered on Instagram, participants frequently discussed other social media platforms (e.g., TikTok, Snapchat, VSCO). During the interview, participants used screen-sharing on Zoom to walk us through the content they shared on their Instagram account (e.g., stories, posts). To note, we only looked at the content the participant posted and not at any posts from people they followed. Per the participant's approval, we also audio-recorded the interview, subsequently anonymizing and transcribing it for analysis. 

The demographics of our interview participants and survey respondents are included in Tab.~\ref{tab:demographics} and Appendix~\ref{sec:survey_results} respectively. In total, we were able to reach participants located in 23 different states through our survey, with the majority of participants concentrated in New Jersey and Florida, and six states in our interview. On average, our participants were $16.4$ years old with ages ranging from 13 to 19. In total, 9 identified as Asian, Pacific Islander, or biracial Asian / White; 6 as White; 3 as Hispanic / Latine or biracial Hispanic / White; and 2 as Black, African-American, or biracial Black / White. Thirteen of our participants identified as woman. All or our participants had Instagram with most (20) having their main Instagram account set as private. A majority (13) had Snapchat; some used TikTok (10), Twitter (8), Facebook (6), and VSCO (4); and a few (<3) used LinkedIn, Pinterest, and BeReal.

\subsection{Analysis}
Grounding ourselves in prior explorations of teenager privacy on social media~\cite{boyd2011social, boyd2013connected, boyd2014s, marwick2014networked}, we were interested in understanding how conceptions of privacy differ on visual content-sharing platforms and how these conceptions have changed over time. We first analyzed our survey data, looking both at summary statistics and disaggregated results by demographics. The results of the survey helped guide question design for our semi-structured interview. 

All interviews were audio-recorded and then transcribed. After the first ten interviews were finished, we transcribed and analyzed data. First, two of the researchers analyzed the transcripts and met to discuss themes. This resulted in a codebook containing 43 codes in total. Examples of codes include ``managing online presence using offline interactions,'' ``concentric levels of privacy within a platform,'' and ``social norms influencing usage patterns.'' The same authors then independently analyzed three interviews and compared our applied codes, calculating inter-rater reliability as the ratio of agreements (i.e., if a selected passage had the same codes applied by both authors) to the total number of codes applied. After achieving an average inter-rater reliability of $0.767$, we individually coded the remaining interviews. We ran a second wave of interviews with eleven participants. The two waves of interviews are the same, except for in the second set we asked additional questions about changes in privacy perception over time. We reached thematic saturation by twelfth interview~\cite{guest2020simple}, using the subsequent nine interviews to corroborate our identified themes.
\footnotesize
\begin{table}[t]
    \centering
    \begin{tabu}{rrrrr}
        \toprule
         Participant ID &  Age & Gender & Race & Account Setting\\ \midrule
         P1 & 16 & Woman & Asian or Pacific Islander & Private\\
         P2 &  15 & Man & White & Private\\
         P3 & 17 & Woman & White& Private\\
         P4 & 14 & Woman & Asian or Pacific Islander, White & Private\\
         P5 & 17 & Woman & White & Private\\
         P6 & 17 & Non-Binary & White & Private\\
         P7 & 16 & Woman & Asian or Pacific Islander & Private\\
         P8 & 17 & Woman & Hispanic or Latine & Public\\
         P9 & 14 & Man & Hispanic or Latine & Private\\
         P10 & 13 & Woman & Asian or Pacific Islander & Private\\
         P11 & 14 & Man & Asian or Pacific Islander, White & Private\\
         P12 & 17 & Man & Asian or Pacific Islander & Private \\
        P13 & 18 & Man & Black or African American, White & Private \\
        P14 & 18 & Man & White & Private \\ 
         P15 & 19 & Man & Hispanic or Latine, White & Private \\
        P16 & 18 & Woman & Asian or Pacific Islander & Private\\     
         P17 & 18 & Woman & Asian or Pacific Islander & Private \\
         P18 & 18 & Woman & White & Private \\ 
         P19 & 17 & Woman & Asian or Pacific Islander & Private \\ 
         P20 & 15 & Woman & Black or African American & Private \\
         P21 & 17 & Woman & Asian or Pacific Islander & Private \\
         \bottomrule
    \end{tabu}
    \caption{Demographic information and identifiers for all interview participants.}
    \label{tab:demographics}
\end{table}
\normalsize
\section{Results}
Following from Marwick and boyd's findings~\cite{marwick2014networked}, we conceptualize privacy as an individual's ability ``to control their situation,
including their environment, how they are perceived, and the information that they share.'' We intentionally adopt this broad interpretation of privacy to allow for our teenage participants to provide their own definition of how they view privacy. From our interviews, we further found that participants interpret privacy not only as who has access to their content but also who can understand what the content means. Further, participants also frequently discussed privacy in terms of the consequences of their privacy decisions, such as their control over how they are being perceived through the content they share or in-person drama that may arise from online interactions.

\subsection{How Do Teenagers Interpret Privacy?} 

\begin{figure}
    \centering
    \includegraphics[width=0.95\linewidth]{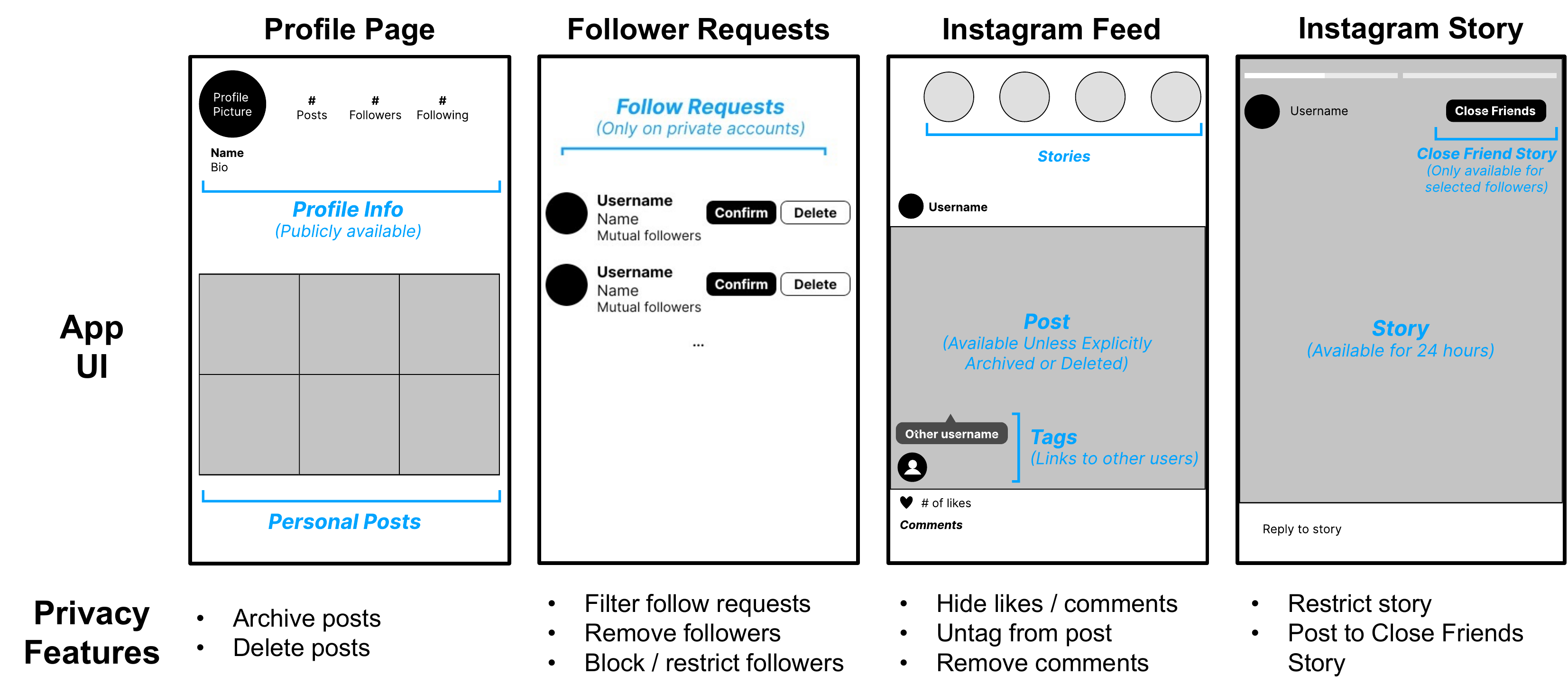}
    \caption{Visualization of the core features and user interface on Instagram, including the profile page, follower requests, feed, and story. For each interface, we also include the built-in privacy features that were most commonly used amongst interview participants.}
    \label{fig:overview}
\end{figure}

Participants interpreted privacy as the ability to control who has access to their content. Forms of control range from filtering follower requests on private accounts to curating content to differing degrees depending on its audience (see Fig.~\ref{fig:overview}). The visual nature of Instagram also introduces privacy concerns regarding posting pictures that reveal faces or location. Additionally, while trust continues to be important in navigating privacy, definitions of digital trust have become more expansive relative to that described by Marwick and boyd~\cite{marwick2014networked}.

\subsubsection{``Do I have mutuals with them?'' -- Trust through association.}
Participants tended to interpret ``privacy'' as their ability to limit access to who can view the content they share on the platform. All except one of the interview participants and $76.4\%$ of survey respondents have private main accounts, using follower requests as their primary form of achieving privacy. They characterized follower requests as a mechanism for filtering out ``random people'' from being able to view their photos. Conversely, their followers are generally people they trust or feel comfortable with seeing the content that they do share. 

Importantly, definitions of digital trust have become more expansive over time. In contrast to Facebook connections which are typically only people the individual knows in person~\cite{dennen2017context}, knowing or having interactions with a potential follower in person is not a prerequisite for trust on Instagram. When survey respondents were asked how many of their 10 most recent followers they interacted with in person, the average response was $4.67$. Similarly, all interview participants had expansive definitions of who they trusted, including peers who went to their school, other high school students from the same area, or people with whom they had mutual connections (``mutuals''). Compared to a Pew Research report from 2013~\cite{pew2013teens} that finds only $33\%$ of teenagers are friends with people they have never met in person, all of our participants had followers they did not know offline. In fact, these ``mutuals'' tended to comprise a large portion of their follower base.  

Contrary to prior works~\cite{may2014safeguarding, badillo2019stranger} that suggest children are easily deceived into trusting unknown individuals online, most participants (N=13) described having concrete heuristics for defining which followers requests are trustworthy when they do not know the individual in person. Namely, ``trust'' on Instagram is judged using both the quantity and quality of mutuals with the requested follower. P2 and P5 stated that if a follower has more mutuals than a pre-defined threshold-- 50 and 80 respectively, they would accept their follow request. For example, upon receiving a follow request, P2 has a defined process for judging trustworthiness: 

\begin{quote}
  \textit{``A really interesting metric is if I see they’re followed by 80 other people that I follow and I'm like `okay, they must be connected.' If it's under five people, I have to vet them, like make sure they're good and make sure I know them somehow, make sure they go to my school or something.''--P2}
\end{quote}
Similarly, P1 checks for the number of mutuals, and if they do not meet her threshold, she will ask the mutual connections she has with the follower \textit{``how do you know them, do you trust them, and do you think I should accept their request''} before deciding on the request. 

\subsubsection{``They know I'm careful'' -- Knowledge of networked privacy.} 
\label{sec:trust}
\begin{figure}
    \centering
    \includegraphics[width=0.75\textwidth]{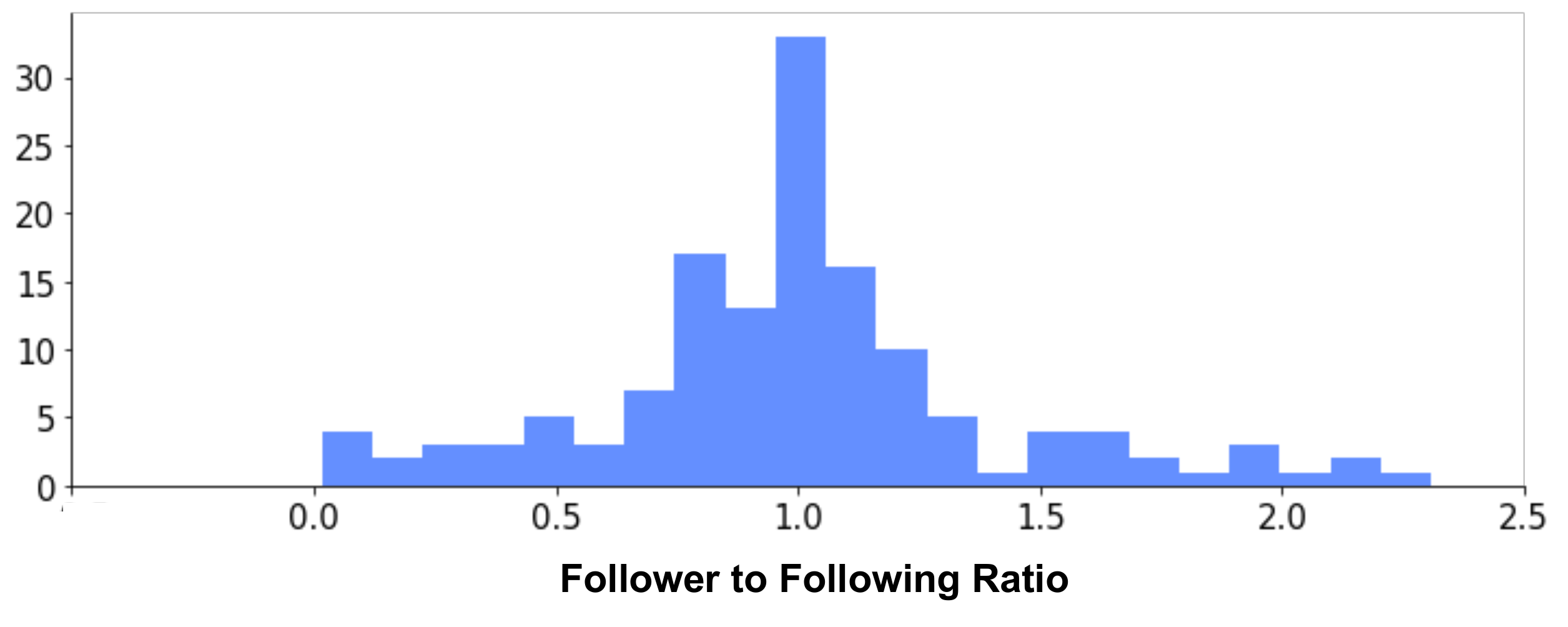}
    \caption{Visualization of the follower to following ratio on the main Instagram account of survey participants. The mean follower to following ratio is 1.98, and the median ratio is 1.01.}
    \label{fig:ratio}
\end{figure}

Trust continues to play an integral role in how teenagers manage privacy in networked settings~\cite{marwick2014networked}. However, while Marwick and boyd~\cite{marwick2014networked} identified ``mutual information sharing'' as an important method for building trust, this is less relevant for our participants. In part, this may be because content-sharing platforms, such as Instagram, allow for asymmetric relationships (i.e., $A$ can follow $B$, but $B$ does not need to follow $A$). This is in contrast to friendship-based social media networks, such as Facebook or Snapchat, on which relationships are inherently mutual. Another motivation for this could be that having more followers can be an indicator of social capital or peer approval~\cite{sciara2021going,chua2016follow}. From our survey, we found the median ratio of followers to following for participants' main accounts is 1.01 and the mean ratio is 1.98 (see Fig.~\ref{fig:ratio}). This indicates that, on average, survey participants are not following all the accounts that follow them, going against the need for reciprocity, found in prior studies~\cite{marwick2014networked,locke2010eavesdropping}, to build trust online.

Since ``mutual information sharing'' appears to no longer be as important of a means for fostering trust online, the teenagers we interviewed have new means for measuring trust. One way participants gauge trust is whether the other person has a similar definition of privacy. Participants are cognizant of the fact that their privacy is linked to the people they are connected with on Instagram. When asked whether she feels comfortable being tagged in posts, P3 responded that she has no concerns with being tagged since \textit{``it's other people who kind of filter their followers the same way.''} P4 reported that her friends allow her to tag them for the same reasons: \textit{``Most of them usually just say yes [to being tagged] because they know I'm careful about who I let follow me and who sees my stuff.''}

We also find that digital trust is persistent, and, at times, can outlast offline trust. For instance, when participants were reviewing their Close Friends list -- a built-in Instagram feature that allows users to share private stories for a more select group of users-- they frequently noted that the list was outdated, and they were no longer friends with some users. While participants detailed updating their Close Friends list to include new relationships they have formed, it appears that they are less likely to remove people. 

\subsubsection{``I don't have any pictures of my face.'' -- Privacy in visual content.} On visual content sharing platform, there are particular aspects unique to images (or videos) that raise privacy concerns. On the most extreme level, P10 detailed that she and her friends do not like posting pictures of themselves; in particular she does not want to post any pictures of her face online. While P4 does post some photos of herself, she chooses to only include images in a group setting, particularly for her profile picture which is publicly accessible. She does this so that when people who do not follow her, namely strangers, view her profile, they will not know who owns the account.

Some participants are also cognizant that sensitive background information can be in their images. Two participants mentioned they do not share images in or of their house. P7 detailed that she tries to generally conceal information that may reveal her location: \textit{``I don't really tend to post pictures of my house or like inside my house, outside my house, around my school, on my story or on my feed.''} When she does post content that might have location-sensitive information, she staggers her post such that she is no longer in the area when she posts: 
\begin{quote}
    \textit{``
    If I'm somewhere and I'm doing something that could say `I'm in New York' and I take a bunch of pictures there, I don't post that on VSCO while I'm in New York. I wait until I get home or somewhere else so because I’m not sure if the location is tied but it's just better to be careful.'' -- P7
    }
\end{quote}

Younger participants, or new users, seemed more likely to express apprehension about sharing their location or personal appearance. P15 shared he used to be concerned about revealing information about his location; however, over time, he realized that \textit{``the chances of [someone finding his location] happening are very low, so [he's] not really nervous about location stuff anymore.''}

\subsubsection{``I felt like I have to be more presentable'' -- Curation as privacy.} Privacy was also equated with the level of curation participants applied to their posted content. This differentiation is applied to what types of content participants feel comfortable posting to their feed instead of their stories. Many participants (N=12) expressed that they only wanted to post to their feed when something important or exciting happens. Common examples that warranted posting include major life events (e.g., graduation, birthdays), vacations, or holidays. On the other hand, participants added content to their stories on a much more frequent basis. In line with prior work~\cite{trieu2020private}, we found that participants shared a wider array of content on their stories (e.g., political infographics, celebrity posts, memes) since they knew the content would only be available for 24 hours. However, participants were more discerning of the stories added to their highlights (i.e., made permanently visible on their profile), keeping only the most \textit{``aesthetic''} (P1) stories to display.

Following the recent research on secondary account usage on Instagram~\cite{dewar2019finsta, taber2020finsta, xiao2020random}, we observed that participants also maintain different levels of formality over different Instagram accounts. For their main accounts, participants recounted feeling the need to \textit{``look more serious and respectable''} (P1) as their parents or other adults would follow them. However, five participants have additional accounts (i.e., ``finstas'' or ``spam'' accounts) that are less curated. For instance, P1 reported that she used her secondary account to express herself more fully compared to her main content which was highly curated: \textit{``I could post silly pictures of something else, or a strange story where I am like `I'm so tired right now' or something more casual that I would like text someone.''} P7 also described filtering the images she posts on her main account to avoid `\textit{`repeating too many of the same poses''} and \textit{``double posting in the same outfit.''} However, she maintains a secondary account, which she refers to as a ``spam'' account, where she can post less curated photo dumps and sillier content.

\subsection{Who are Teenagers Getting Privacy From?} 
Participants had a clear sense of who they were seeking privacy from. For many, the privacy threat was direct: parents or authority figures who followed the participants on Instagram. However, imagined or indirect threats were often a catalyst for teenagers to implement privacy mechanisms. 

\subsubsection{``Sometimes I block my family members'' -- Managing direct supervision.} 
\label{sec:parents}
The most common people that participants tried to restrict on their accounts were adults. Many participants (N=8) were followed by their parents or other adult authority figures (e.g., relatives, teachers). Unlike strangers whose requests can be filtered out on private accounts or deleted on public accounts, participants had close relationships with these individuals and had to give them content access privileges. P3 felt obligated to allow adult family members to follow her account, even though their presence limits the type of content she feels comfortable sharing. 

We find that participants have differing perspectives on adult supervision. On the one hand, P7 likes that her mother follows her main and secondary Instagram accounts, even though her mother has restricted her from sharing certain content, such as bikini pictures. In fact, P7 described using her mother as ``\textit{an extra security standpoint}'' for determining what content was appropriate to post. However, other participants viewed direct supervision as less favorable. Most (N=5) employ a variety of privacy features to restrict adults from certain parts of their online presence. Some techniques include blocking them from stories or creating secondary ``finsta'' or ``spam'' accounts that adults do not know about. Even for P7, she does caveat that she restricts her mother from viewing the stories on her spam account using the ``block'' feature, as she tends to share images of herself crying. 

In line with prior work~\cite{dennen2017context}, we find that all of our teenagers who have adult followers anticipated context collapse~\cite{marwick2011tweet} and implemented privacy measures to avoid this phenomenon. While participants are comfortable with adult followers seeing the posts on their main accounts, which they deem as more ``\textit{palatable}'' (P3), some (N=5) restrict adults' access to their stories or secondary accounts. Although participants maintain that they do not post anything ``\textit{incriminating or illicit}'' (P3), they tend to feel uncomfortable knowing that adults can view their stories. P3 described her rationale for blocking older family members from her story:
\begin{quote}
    \textit{
    ``Pretty much anything I put on my story is something I don't want [adults] to see because again I feel they're kind of out of touch. It's not anything bad that they would be like `why are you posting that.' But they don't listen to the music that I listen to and I feel like we're from a different generation. There's a communication gap there.'' -- P3
    }
\end{quote}

\subsubsection{``They were a little creepy'' -- Managing strangers.} 
Following the threat of direct supervision, participants were most concerned about strangers viewing their content. This is unsurprising given the attention that educational programs and parental concerns place on external threats, such as ``stranger danger'' on online platforms~\cite{livingstone2014developing, badillo2019stranger, zhang2016nosy, boyd2013connected}. P4 described how her parents warned her about strangers online: ``\textit{My parents, they're not a very strict pair of people but they've definitely taught me that there are a lot of weird people out there.}'' P3 also expressed a fear of being preyed upon or being stalked via her social media accounts.
 
When discussing why they chose to have private accounts, most participants (N=14) mentioned that they wanted to filter out ``random people'' from following them. Unlike their followers, who participants described as people they trust, the rejected followers were characterized as ``creepy'' or ``weird.'' Eleven participants mentioned having to filter out bot accounts that tried to follow or message them, which often included inappropriate or sexual content.

In contrast to prior work that found adolescents were not adept at identifying the age of strangers online~\cite{badillo2019stranger, may2014safeguarding}, participants were cognizant when unknown followers appeared to be older than them. P9 recounted that he used to have a public Instagram account. However, he felt uncomfortable and later changed his account to private after observing that many of the followers he did not know personally were older. For P8, our only participant with a public main account, she described how older men would harass her, leaving inappropriate comments or tagging her in unrelated, explicit posts.

\subsubsection{``I don't want to give them fodder for gossip'' -- Managing in-person drama.} Similar to previous findings~\cite{marwick2014networked}, in-person drama continues to motivate the use of different privacy features. Some participants (N=5) described using privacy mechanisms to avoid potential drama that may occur. For P3, this stems from a general sense that teenagers \textit{``are just very gossipy.''} As a result, she does not accept ``follow'' requests from people who do not like her as a means of avoiding potential harassment. Sometimes, the dissemination of drama online can be systematic, as participants mentioned knowledge of anonymous gossip accounts run by their peers. 

In addition to general concerns about in-person drama, participants also implement temporary privacy features in response to specific events. P7 described she and her ex-boyfriend mutually restricted each other following their break-up. Restricted accounts still show up as followers; however, if her ex searched her account, he would be shown an error. This privacy mechanism was a strategic maneuver to avoid potential drama since she still had to interact with her ex in-person: \textit{``I'm like `okay, we need space,' but I don't want to start any drama because we had to work together and I didn't want anyone to spread ridiculous rumors.''} Overall, participants were concerned that their Instagram activity would lead to negative in-person interactions. 

Conversely, some participants (N=4) also enacted privacy features as a reaction to in-person drama. A common technique that participants use is removing content related to the drama, such as deleting posts or untagging themselves in images that others have shared. Participants expressed that untagging themselves from images with them and that person allowed them to sever connections with those they \textit{``didn't want to be associated with anymore''} (P1). Further, they could do so without having to directly ask people to delete the post. Being able to act discreetly is crucial as not to exacerbate existing antagonistic relationships. P7 stated that before removing an unwanted follower \textit{``refrains from posting for a while, so they wouldn't see me on their feed''} to avoid additional conflict. While using privacy features could potentially foster more gossip, none of our participants reported that their actions on Instagram translated into in-person confrontation.

\subsection{How do Teenagers Navigate and Configure Privacy?} 
While participants had methods of limiting access to meaning, they generally used the built-in Close Friends list and self-configured secondary accounts to limit access to content. These features facilitate the creation of concentric content-sharing circles, allowing different subsets of followers to see specific content. Participants also consider content and behavior across platforms, treating their various online accounts as interconnected.

\subsubsection{``Obviously I asked if I could post'' -- Online and offline methods for managing privacy.} 
First, we explore what built-in privacy features participants use. All participants mentioned making use of Instagram's built-in privacy functionality. The most common is, of course, creating a private account and filtering follower requests. Nonetheless, participants utilize additional privacy features within their private accounts. The most common features related to content management (e.g., creating ephemeral stories, archiving posts). Other popular features centered around limiting follower access (e.g., blocking users from stories, using the Close Friends list).

In addition to using Instagram's built-in features, participants configured their own privacy practices. Similar to prior works~\cite{xiao2020random, dewar2019finsta, taber2020finsta}, we found a rich use of secondary accounts (``finsta'' or ``spam'' accounts). Outside of secondary accounts, two participants mentioned using bookmarking as privacy tools: P3 bookmarks posts she enjoys but may be controversial instead of liking the post and P17 bookmarks public posts she comments on so she can later delete her comments.

Finally, we find that technical features are not always sufficient. In line with Marwick and boyd's findings~\cite{marwick2014networked}, interpersonal relationship management play an integral role in maintaining privacy online. Participants frequently mentioned using offline interaction to determine their online presence. In contrast to prior findings~\cite{rashidi2020s} that examined how offline interactions are used to manage undesirable content reactively, we find that participants are using in-person interactions to \textit{proactively} manage content. This is best illustrated through the processes that survey respondents use before posting content or tagging friends in posts. $68.9\%$ and $33.0\%$ of survey respondents stated they asked their friends, usually via text or in-person, for permission before posting content (i.e., post or story) and tagging in a post respectively. As navigating privacy in networked settings continues to be a challenge technically, offline social mechanisms used to regulate online presence may become more of a necessity. 

\footnotesize
\begin{table}[]
    \centering
    \begin{tabular}{lcccccccccccccccccccccc}
    \toprule
         &  1 & 2 & 3 & 4 & 5 & 6 & 7 & 8 & 9 & 10 & 11 & 12 & 13 & 14 & 15 & 16 & 17 & 18 & 19 & 20 & 21\\
         \midrule
         Close Friends list &  $\circ$ & $\bullet$ &  & $\bullet$ & $\circ$ & $\circ$ & $\bullet$ & $\bullet$ & $\bullet$ & & & & $\bullet$ & $\circ$ & $\bullet$ & $\circ$ & $\circ$ & & & $\bullet$& $\bullet$\\
         
         Secondary accounts & $\circ$ & & $\bullet$ & & $\circ$ && $\bullet$ & $\bullet$ & & & & & & & $\circ$ & $\bullet$ & $\circ$ & $\circ$ & & $\bullet$& $\circ$\\
         
         Snap private stories & $\bullet$ & & $\bullet$ & $\bullet$ & $\bullet$ & $\bullet$ & & & & & & & $\bullet$ & $\bullet$ & $\bullet$ & $\bullet$ & & $\circ$ & & & $\bullet$\\
    \bottomrule
    \end{tabular}
    \caption{Comparison of three commonly mentioned features and practices for limiting content-sharing circles. The $\bullet$ indicates that the participant currently uses the feature. The $\circ$ indicates that the participant has used the feature before but either did not do it frequently or no longer uses it. In addition to follower curation, we see that most participants have used or currently use two different methods for limiting content.}
    \label{tab:features}
\end{table}
\normalsize 

\subsubsection{``It's a much smaller circle'' -- Concentric circles of trust.} 
\label{sec:concentric}
We further explore the two most common methods for limiting access to content: the built-in Close Friends list and self-configured secondary accounts. During our discussion of these features, we found that participants would frequently mention their use of Snapchat private stories either in conjunction or in place of the features on Instagram. All three features are used to create more constrained content-sharing circles. When describing what they shared, participants characterized the content as being more authentic. They were more willing to share ``silly'' or ``random'' stories as well as more emotional content, such as images of themselves crying or ``ranty captions.'' Participants felt comfortable sharing this content with their audiences both because they trusted these followers more and because they felt these followers would find the content relevant.

As shown in Tab.~\ref{tab:features}, while most participants make use of either Close Friends list or a secondary account, seven of our participants have tried both features, and three actively use both. For P8, she considers her secondary account followers as a subset of her Close Friends, which are in turn a subset of her followers on her main account. P8 described how she delineates who has access to each account: ``My Close Friends is basically all the people that would be in my finsta and people that I trust, but not enough to be on my finsta.'' Similarly, P7 expressed a similar phenomenon in which some of the followers on her Close Friends are not allowed to follow her secondary account since she ``wouldn't feel as comfortable with them seeing content I posted on my second account.'' She also noted that, while she does use both, she has started to use her secondary account as a replacement for her Close Friends list. This sentiment is echoed by other participants who tended to use only one of these privacy mechanisms.

Even though the technical offerings for Instagram's Close Friends and Snapchat's private stories (e.g., available to select followers, ephemeral) are similar, participants had a clear delineation between the two, preferring to share more personal and intimate content to Snapchat. Participants gave two main reasons for this. First, some participants (N=4) found Instagram's story interface to be cumbersome with too many features whereas Snapchat was simpler. Second, even though Close Friends was only available to a limited audience, participants still associated it with their main Instagram account and felt the need to present a more curated or polished representation of themselves.

\begin{figure}
    \centering
    \includegraphics[width=0.35\textwidth]{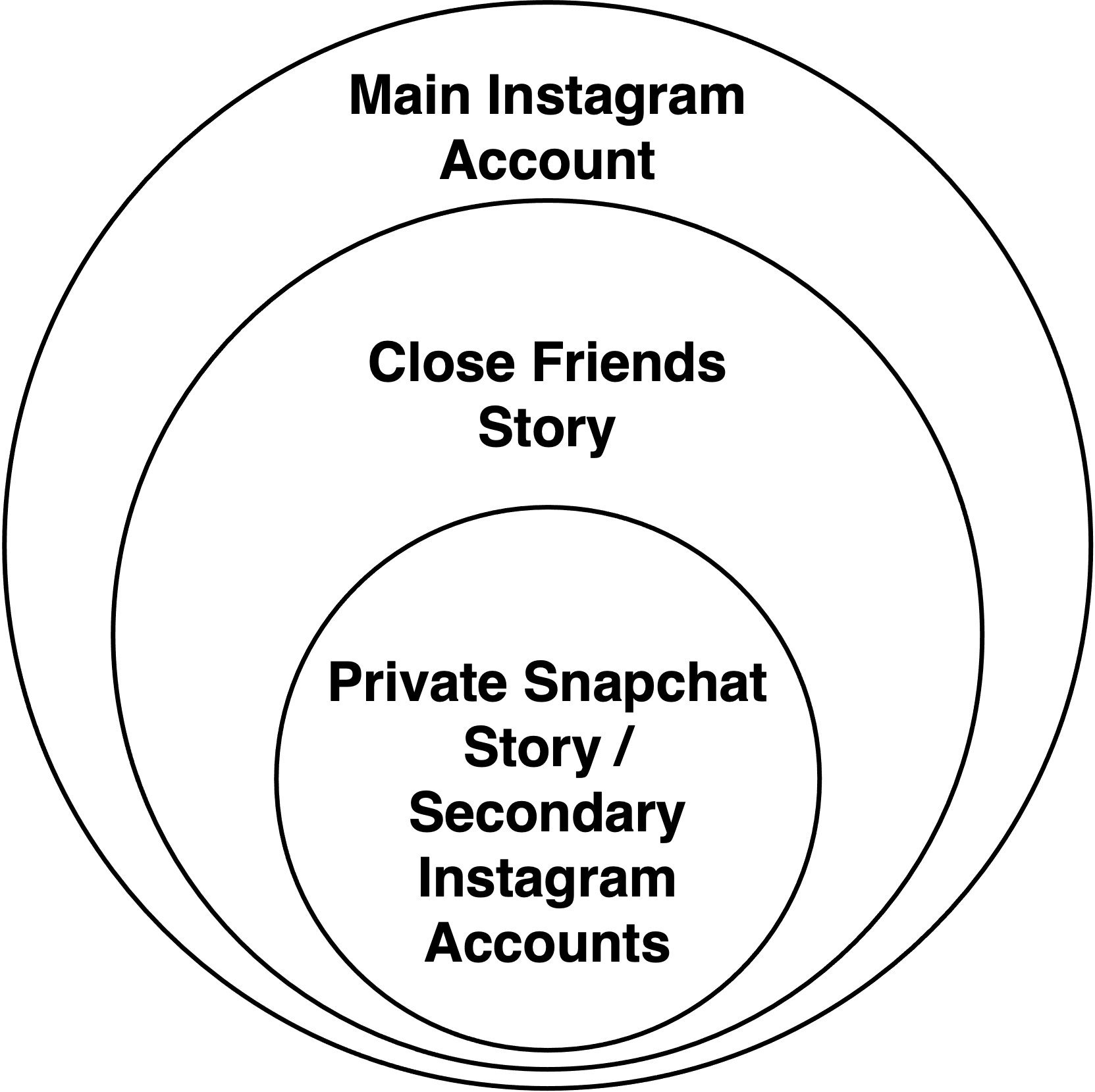}
    \caption{Visualization of the concentric circles of trust that participants create on Instagram.}
    \label{fig:concentric}
\end{figure}

\subsubsection{``A subtle hint'' -- Limit access through meaning.} 
\label{sec:steganography}
While a large focus lies in limiting access to shared content, participants configure privacy by limiting access to meaning. As discussed in previous works~\cite{marwick2014networked, oolo2013performing}, teenagers continue to utilize social steganography or ``hiding content in plain sight'' on Instagram. P2 described that he and his friends included fruit emojis in their bios that have the color of the bisexual flag to signal their sexuality. He expected other teenagers would understand how to interpret the message in the emojis. However, emojis are common enough that people who do not understand the meaning would not react negatively. 

\subsubsection{``It's just a link to my VSCO'' -- Privacy across platforms.}
Finally, we examine how privacy on Instagram is negotiated within a larger ecosystem of social media platforms. Similar to most US social media users~\cite{pew2015teens}, participants have social media accounts on multiple platforms and must navigate privacy networked not only within Instagram but also across other social media platforms. The most commonly mentioned platforms include Snapchat, TikTok, and VSCO. 

\begin{figure}
    \centering
    \includegraphics[width=0.8\textwidth]{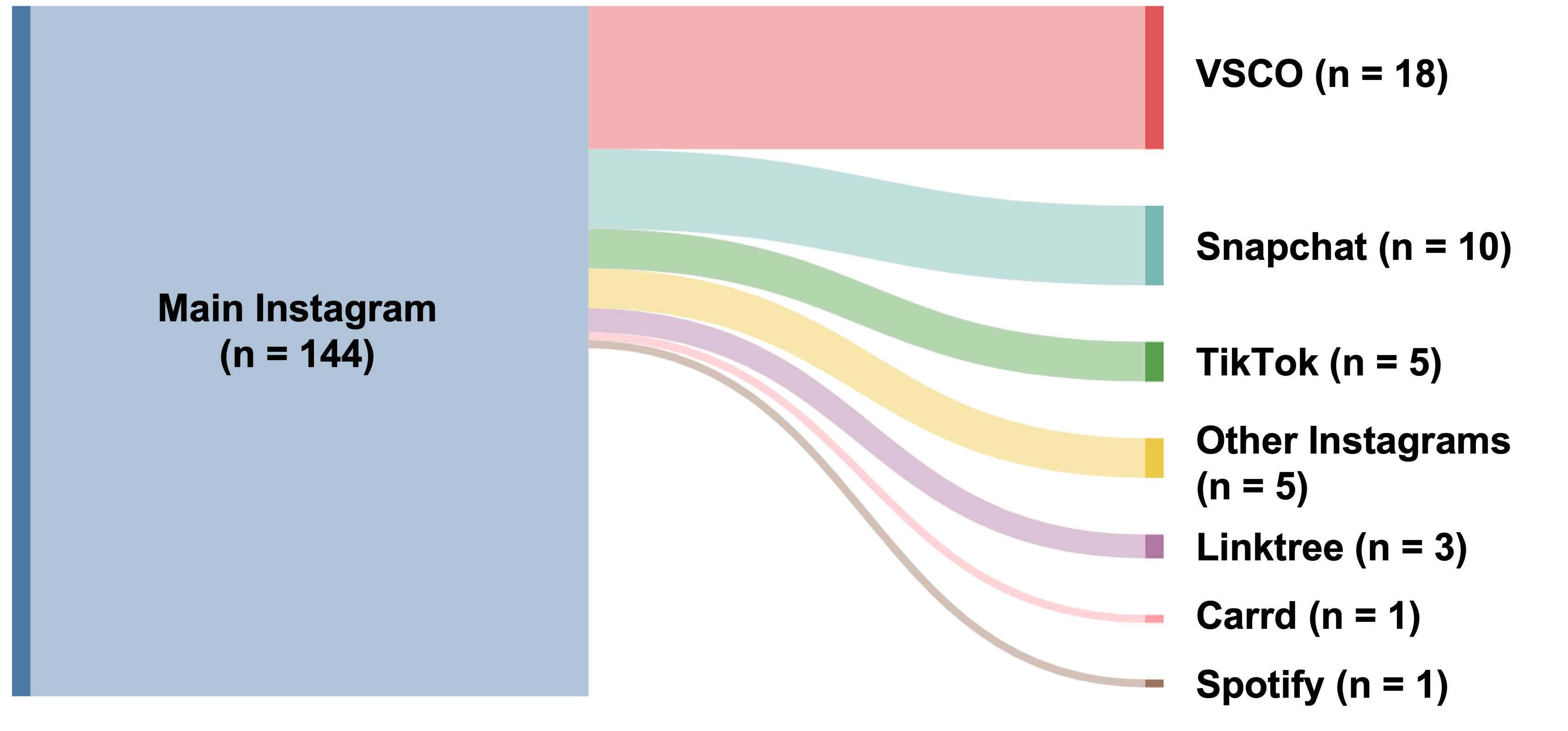}
    \caption{Visualization of the most common links to other social media platforms or accounts that survey participants include in their bio. $29.9\%$ (N=43) of participants include a link in their bio, with the most common links being to VSCO and Snapchat.}
    \label{fig:network}
\end{figure}

Compared to a 2016 study on how users navigate multiple social platforms~\cite{zhao2016social}, we find participants are more inclined to create permeability across social media platforms. Rather than viewing platforms as separate entities, participants perceive different platforms as inherently connected, in what Devito et al.~\cite{devito2018too} refer to as a ``personal social media ecosystem.'' This is best illustrated through the explicit linking or references to different platforms. As shown in Fig.~\ref{fig:network}, $29.9\%$ of survey participants and five interview participants include links (e.g., hyperlinks, username handles) in their Instagram to other platforms. Thus, these teenagers must navigate privacy boundaries not only within but across platforms. For example, P7 has a private TikTok with a smaller subset of followers compared to her main Instagram. She will occasionally repost her TikToks onto her Instagram story if they contain \textit{``content [she's] usually fine with other people seeing.''} 

There is a distinction between external social media accounts that are more public-facing and private compared to Instagram. This delineation is influenced by both the technical privacy features and the culture on these platforms. For example, three participants link their VSCO \footnote{https://vsco.co/}, an alternative photo-sharing platform, from their Instagram bio. Compared to Instagram, VSCO has a narrower set of privacy options~\cite{vsco} as all accounts are public. Even though two of the three have private Instagrams, they share content publicly on VSCO more frequently than on Instagram. However, since VSCO is a more public platform, P7 acknowledged that more people would look at her VSCO content and is more careful about what she posts on VSCO. Similarly, on TikTok, P1 and P5 reported having public accounts even though their Instagram accounts were private. They both considered TikTok a platform for consuming content, whereas Instagram is also for creating content. 

While TikTok and VSCO tend to be more public forms of social media, participants use Snapchat, particularly private stories, to share more personal content. In part, this difference in usage also stems from how participants view the platform's purpose. Whereas Instagram is described as being for a broader audience and disseminating information, participants only used Snapchat as a means for connecting to people with whom they have closer relationships. P4 provided her rationale for sharing more private information on Snapchat: 
\begin{quote}
    \textit{``Snapchat can be as private as you want it to be without the settings being changed because your Snapchat is essentially like a private Instagram account because you get to see who wants to be your friend and you don't necessarily have to snap them. You could just be friends with them, whereas on Instagram, once they follow you they can see everything, it’s like you have no control.'' -- P4
    }
\end{quote}

\subsection{Why Do Teenagers Make These Decisions about Privacy?} 
Participants mainly made decisions with others in mind, both their followers and an imagined audience. Oftentimes, participants wanted to broadcast information to their followers, whether to connect and engage, or to spread awareness. They were also aware of how others use Instagram and its privacy settings.

\subsubsection{``Just scared of being judged'' -- Importance of social perception.}
\label{sec:judged} 
Social expectations played a large role in how participants navigate their activity on Instagram. While some participants (N=6) were adamant that social validation was not a reason as to why they used Instagram, participants also expressed that they used the platform or chose to post certain content since it was expected of them. Some participants (N=7) also described feeling pressured to initially make an Instagram account to fit in with their peers. At the most extreme, P17 stated her friends set up an Instagram account for her so that she could follow their accounts and would pressure her to post on the platform. Similarly, many participants stated they felt obligated or were pressured by peers to post, especially during significant events (e.g., graduation, birthdays, college commitment). Participants expressed this was because they noticed their peers posting these moments and wanted to show their followers they were also having fun.

These social expectations also influence how teenagers configure their privacy settings. This is reflected in how teenagers manage their follow requests. In Sec.~\ref{sec:trust}, we discuss how teenagers tend to have more followers than those they are following; this is in part motivated by social perception. P16 explained that she used to care about her following to follower ratio since it made her appear \textit{``cooler''} to have more people interested in her life. Similarly, P13 reflected that he used Instagram metrics, such as followers or likes, to gauge how well he was \textit{``integrating socially''} at his new school.

Social validation cues, such as follower counts, influenced who participants let access their content. For example, P14 engaged in more privacy risky behavior, allowing bot accounts to follow him but not following them back to increase his follower count. On the other hand, participants would limited their followers to increase their follower to following ratio. P15 and P21 both mentioned that they would unfollow someone if they noticed that user was not following them back. In fact, P15 even used a third party application that tracked which of his followers did not follow him back. 

Participants also expressed being aware of how others perceived their actions on the platform. Almost all participants (N=14) mentioned they preferred sending funny public posts via direct message (DMs) rather than tagging their friends, as they only wanted this content to be seen by the individual with whom they were sharing the post. Participants had a strong desire to interact with public posts but did not want to be noticed by others excluding their select group of friends. P1 explained that the content she DMs typically is embarrassing or related to something she did not want other people to know about. P3 had a similar rationale and noted that she tended to privately share posts that were \textit{``silly''} or that she \textit{``did not want to be associated with in real life.''}

\subsubsection{``That's what everybody does'' -- Learning privacy configurations from peers}  
\label{sec:peer}

Participants are aware of how their peers use Instagram, learning about and configuring their privacy settings based on their peers. We found that survey respondents learned about different privacy features via word of mouth or observing other people's accounts $24.4\%$ of the time. Knowledge of certain features, such as Close Friends Stories, are even more likely ($48.6\%$ of the time) to be be spread through these peer-to-peer pathways (see Appendix~\ref{sec:survey_results} for more details). For example, while P21 has never personally made a story before, she is aware of the Close Friends feature as she has seen the stories her friends' post.

Alternatively, teenagers use their peers as examples of what \textit{not} to do on the platform. P4 explained that her decision to remain private on the platform was in part motivated by the comments or messages her friends received: \textit{``A lot of my friends have public Instagrams and they aren't as careful about it as someone I think should be, especially about Instagram. They'll just kind of post very interesting posts, but they'll tell me things like, `Oh this random person went into my DMs and they sent some really inappropriate stuff.'''}

\subsubsection{``I didn't like how I looked'' -- Body image concerns}
\label{sec:body}
Due to Instagram's visual nature, participants were conscious of how others may perceive their appearance and used privacy features in reaction to judgement they faced. This notion was especially apparent for our female participants. Nine of the thirteen teenage girls interviewed mentioned using privacy features due to appearance or body image issues. For these participants, concerns about body image led to self-censorship. For example, a common reason participants gave for untagging or archiving posts was that they did not look good in the image. This sentiment is mirrored in our survey results. Of the participants who have untagged or archived posts, $43.5\%$ and $30.0\%$ respectively cited not liking the way they look as the reason why (see Appendix~\ref{sec:survey_results}).

One participants (P8) experienced direct harassment in the comment sections of her posts. She described that some of these comments were related to her appearance: \textit{``I've received comments of the type `oh, you forgot to suck your belly' and `you're looking fat,' that kind of stuff.''} Most of these comments about body image tend to come from other women; however, comments or tags related to inappropriate, sexual content tend to come from men. In turn, P8 used her private account to share content that she felt uncomfortable having on her public account. When detailing what content she posts to her private account, P8 explained, \textit{``I'm just scared of being judged in most of those videos that I put on my finsta because probably I’m without makeup or without my hair done and that type of stuff.''}

\subsubsection{``Clearly done by a younger me'' -- Changes in privacy perceptions} 
\label{sec:time}

Finally, teenager perceptions of privacy are not static. We found not only that many participants (N=9) observed that their own privacy configurations had changed since starting their account, and, as a result, the privacy perceptions of our younger participants often differed largely from older participants. Many participants (N=10) described a similar trend of the content they posted on Instagram: when they first got the account, they often would post \textit{``random''} content such as blurry photos or scenery images. Most of this content eventually was deleted or archived for being too \textit{``cringy''} or \textit{``embarrassing.''} As participants grew older--often coinciding with entering high school, they felt a need to create a more curated presence, posting only during important events. Finally, veteran users, typically those ages 17-19, reported that they were more willing to post more casual photos. However, these were posts are still different than the random posts that younger teens share; instead, participants described detailed thought processes that went into selecting what they shared, representing a more casual and personal but still highly curated version of themselves that they choose to share. 
    
More experience on the platform also coincides with participants having laxer privacy perceptions. As discussed in Sec.~\ref{sec:trust}, older participants were more likely to share information about their location. Participants rationalize that they have not experienced any negative repercussions when sharing more personal information. Further, while younger participants had more stringent criteria before accepting a follower they did not know in person, most of the older participants (N=6) reported they would be willing to accept anyone who went to their school even if they did not have mutual connections. However, while participants grow more open about what they share, they also develop more complex configurations to manage privacy. Older participants were more likely to have used both technical features for managing who views their content (e.g., Close Friends list, blocking / restricting followers) and making use of external platforms like Snapchat to share more intimate posts. This mirrors Adorjan and Ricciardelli's~\cite{adorjan2019new} ``new privacy paradox,'' which states that teens claim to ``have nothing to hide'' while still using complex privacy management configuration.
\section{Discussion}
    \subsection{Changing privacy perceptions over time} 
    Teenagers' understandings of privacy are not monolithic and are malleable to change over time. As our findings show (Sec.~\ref{sec:time}), participants' perceptions of privacy change as they grow older. This is reflected in the content they share, the people they trust, and the features that they use. While we cannot disentangle these changes from shifts in platform culture over time, we do observe differences in privacy perceptions between our younger and older participants. Prior work~\cite{livingstone2008taking, xie2015see} has also noted that age influences privacy perceptions and configurations across other platforms including Facebook and MySpace. 
    
    Despite these differences, many social media platforms take a one-size-fits-all approach towards teenage safety, ignoring the nuances of how teenagers interact with these platforms. A better alternative is to segment the teenager population and offer different safety options for different age brackets. TikTok has already started to implement this approach. For example, downloading videos created by users under the age of 16 are disabled but enabled for teenagers ages 16-17 although the feature is defaulted to off~\cite{tiktok}. In general, privacy features should take into account how adept teenagers may be at navigating at the platform and their level of privacy concern. New users who are unfamiliar with the platform should have the strictest safety options. As users grow older and become more familiar with the platform, the safety features should adapt to these change, although the option to have more stringent safety guards should, of course, be available. 
    
    \subsection{Networking privacy \textit{across} platforms} 
    In our study, we originally intended to study only Instagram; however, through our interviews, participants would mention other social media platforms they use to configure privacy. Thus, we argue that privacy is networked not only within but across multiple social media platforms. These social media accounts do not exist in isolation but rather play different roles in helping form an individual's digital identity. This is most clearly demonstrated through the bidirectional links participants place in their social media accounts with the hopes that followers from one platform will follow to the other. Thus, while privacy on social media has typically been studied with respect to one platform, our findings suggest that looking at cross-platform dynamics offers a more holistic view of teenager privacy.  
    
    The affordances different platforms offer influence how users manage their privacy settings. Prior works have already analyzed how both low-level (e.g., technical and design features) and high-level (e.g., platform culture) affordances affect self-presentation~\cite{scolere2018constructing, devito2017platforms, taber2018personality}. We find that platform affordances also dictate privacy management. For example, we found that our participants may engage in riskier behaviors on Instagram, such as allowing more people to follow them than those who follow them, since the platform allows for asymmetrical relationships. This is in contrast to platforms such as Snapchat that allow for predominantly symmetrical relationships, and participants are more curated about who they are friends with on the platform. Further, although the Close Friends story and Snapchat private story are functionally the same, participants tended to share more personal information on their Snapchat and to a more select audience. As one participant explained, Snapchat is a platform ``\textit{for friends}'' (P21) whereas Instagram is for content creation and influencing. For designers, this suggests that regardless of what technical affordances a platform may offer, the success and ultimate use of these features is highly tied to the platform's culture.

    \subsection{Designing platforms for teenage safety}
    Finally, we provide a set of design implications that we believe can help platform designers improve teenagers' safety on social media. 
    
    \subsubsection{Empowering, not restricting, teenagers} 
    While there have been efforts from social media platforms, including Instagram, to integrate teenage safety features, most of these efforts have focused on restricting teen usage. For example, in their most recent update\footnote{Released Dec. 2021} on teenage safety, Instagram has focused on launching more supervision tools for parents and guardians. Other features include preventing adults who do not follow teens from messaging, tagging, or mentioning them. In general, safety features have focused on restricting the ways teens use the platform as opposed to giving them more options for controlling their experience. 
    
    Efforts to restrict teenager behavior are both unlikely to be successful and unnecessary. For example, teenagers, especially those that are older than 16, are indifferent or unreceptive to parental supervision~\cite{tufekci2008can}. In fact, parents are sometimes the people teenagers are trying to avoid on social media (Sec.~\ref{sec:parents}). Thus, parental supervision tools, while well-intentioned, can end up being perceived as a violation of privacy~\cite{marwick2014networked}. Ultimately, these restriction are underestimating teenagers' ability to manage their own privacy. In contrast to beliefs that teenagers do not care about privacy on social media~\cite{barnes2006privacy}, teens are generally adept at navigating and configuring privacy in these networked settings. Rather than regulation, we propose that platforms build features that empower and inform teenagers' choices. For instance, instead of restricting all messages from adult accounts who do not follow teens, it would be more helpful to notify the teenager that the user is an adult or even a bot. The platform should give the teenager the agency to delete, block, or reply to the account. 
    
    \subsubsection{Offering more granularity when limiting audiences}
    An important way that teenagers conceptualize privacy is via \textit{who} can view their content. Many social media platforms, including Instagram, use an ``access-control list'' approach to privacy. However, when this control is only applied at the account-level, multiple audiences are collapsed into one group, leading to the potential of context collapse~\cite{marwick2011tweet, dennen2017context}. One way to avoid this is to allow more fine-grained control on individual pieces of content. 
    
    On Instagram, there has already been a trend towards more granular control, as evidenced by the Close Friends story. We advocate that social media platforms give user's greater flexibility when defining access control lists, accounting for complex social relationships that users may want to mirror online. Particularly, for Instagram, participants found the Close Friends feature to be rigid as they were only able to define one set of users as their ``close friends.'' However, participants often had different friend groups with whom they want to share different content. Having to continuously update their Close Friends list can be too cumbersome. In general, we suggest that social media platforms allow for multiple smaller content-sharing circles. Already this is adopted by other social platforms, such as Facebook and Livejournal~\cite{marwick2014networked}. For example, on Facebook, a user can choose to selectively share information with pre-defined groups, such as ``close friends'' and ``coworkers'' which more closely mirror her existing relationships~\cite{marwick2014networked}. 
    
    \subsubsection{Allowing for anonymity} 
    To exist on social media, users have to ``type themselves into being''~\cite{marwick2014networked,sunden2003material} --  a concept that makes teenagers uncomfortable. Content-sharing platforms necessitate that users share information about themselves~\cite{john2013sharing}. For posts and stories on their account, teenagers can rely on access-control-list methods for configuring privacy either by having a private account or using other technical features. This configuration fails when interacting (e.g., liking, commenting) on other people's content. In particular, teenagers expressed discomfort with the idea of other people, especially strangers, being aware of what they are doing online. For example, only one of our participants was willing to comment on public posts; she made the effort to delete all of her comments on public posts after another user liked or replied to her comment. 
    
    Yet teenagers still express a desire to interact with public content. This is evidenced by the fact that they are willing to take the additional step of messaging the post to a smaller circle of friends rather than commenting (Sec.~\ref{sec:judged}). While anonymity is a potential option, it also engenders the potential for an increase in harmful and abusive behaviors~\cite{ma2016anonymity,santana2014virtuous,barlett2016predicting}. Alternatively, platforms could allow likes and comments on public content to be limited between just the user and the content creator. Designers could also extend the access-control-list methods, allowing users to define who can see their likes and comments. 
    
    \subsubsection{Leveraging peer-to-peer pathways for privacy awareness} 
    \label{sec:disc_peer}
    One of our goals is to encourage teenagers to be more aware of built-in privacy features. Despite being regular users of Instagram, participants often confessed to not knowing about many of the features available to them. However, we identified that one fruitful way for learning about privacy features is through peers' accounts or usages. Given the influence of social norms on Instagram, teenagers are more aware of how their peers use the platform, including how they configure privacy, and are influenced by these norms to use certain settings themselves. A study on Facebook~\cite{mendel2017susceptibility} found that users are more likely to be socially influenced by peers than organization when setting privacy features. This effect was more pronounced amongst younger users. Thus, we envision that platforms can leverage the power of social norms with information about privacy features spreading via peer-to-peer pathways. For example, a participant could get a notification that a number of their followers have used the archive feature and provide the option to learn more about the feature. While we acknowledge that awareness of privacy features does not necessarily translate to usage~\cite{hargittai2010facebook}, learning from peers can increase the chance that teenagers will adopt these features~\cite{mendel2017susceptibility,emami2018influence}. 
    
\section{Limitations}
Our participants were all US teenagers, the majority of whom were based in the greater New York City and Philadelphia area. Further, our sample of participants is not representative in terms of race or socioeconomic status. It is possible that a more diverse participant sample may reveal different privacy perceptions and subsequently different management techniques. In future work we will study a larger sample size of participants and use quantitative methods to learn how applicable our findings are to teenagers more broadly.
\section{Conclusion}
In this work, through semi-structured interviews, we demonstrate how teenagers limit access to meaning and content. Furthermore, we explore more deeply into \textit{why} these decisions were made. Participants mainly limited access to content using Instagram's built-in Close Friends list and self-configured secondary accounts. Others, both followers and imagined audiences, influenced their privacy-related decision making. Based on our findings, we make a series of recommendations on how to improve teenagers' control over their privacy on social media platforms.

\section{Acknowledgements}
We would like to thank Janet Vertesi and our classmates in COS 597I: Social Computing for their helpful comments, and Rafael Silva, John Van Horn, Jessica Liu, Sharon Zhang, Jenny Ma, Nicole Meister, Daniel Blackmon, and Peter Horn for helping recruit participants. Finally, we are grateful to our interview and survey participants for their time and insight. 

\bibliographystyle{acm}
\typeout{}
\bibliography{main}

\begin{thebibliography}{10}

\bibitem{vsco}
Private profiles on vscp.
\newblock
  \url{https://support.vsco.co/hc/en-us/articles/360040916071-Private-profiles-on-VSCO}.

\bibitem{acquisti2006imagined}
{\sc Acquisti, A., and Gross, R.}
\newblock Imagined communities: Awareness, information sharing, and privacy on
  the facebook.
\newblock In {\em Privacy Enhancing Technologies Workshop (PET)\/} (2006).

\bibitem{adorjan2019new}
{\sc Adorjan, M., and Ricciardelli, R.}
\newblock A new privacy paradox? youth agentic practices of privacy management
  despite “nothing to hide” online.
\newblock {\em Canadian Review of Sociology 56}, 1 (2019), 8--29.

\bibitem{agosto2017don}
{\sc Agosto, D.~E., and Abbas, J.}
\newblock “don’t be dumb—that’s the rule i try to live by”: A closer
  look at older teens’ online privacy and safety attitudes.
\newblock {\em New Media \& Society 19}, 3 (2017), 347--365.

\bibitem{pew2018}
{\sc Anderson, M.}
\newblock A majority of teens have experienced some form of cyberbullying.

\bibitem{pew2018teens}
{\sc Anderson, M., and Jiang, J.}
\newblock Teens, social media and technology overview 2018.
\newblock Tech. rep., Pew Research Center, 2018.

\bibitem{badillo2019stranger}
{\sc Badillo-Urquiola, K., Smriti, D., McNally, B., Golub, E., Bonsignore, E.,
  and Wisniewski, P.~J.}
\newblock Stranger danger! social media app features co-designed with children
  to keep them safe online.
\newblock In {\em ACM International Conference on Interaction Design and
  Children\/} (2019).

\bibitem{balleys2017being}
{\sc Balleys, C., and Coll, S.}
\newblock Being publicly intimate: teenagers managing online privacy.
\newblock {\em Media, Culture \& Society 39}, 6 (2017), 885--901.

\bibitem{barlett2016predicting}
{\sc Barlett, C.~P., Gentile, D.~A., and Chew, C.}
\newblock Predicting cyberbullying from anonymity.
\newblock {\em Psychology of Popular Media Culture 5}, 2 (2016), 171.

\bibitem{barnes2006privacy}
{\sc Barnes, S.~B.}
\newblock A privacy paradox: Social networking in the united states.
\newblock {\em First Monday\/} (2006).

\bibitem{boyd2010social}
{\sc Boyd, D.}
\newblock Social network sites as networked publics: Affordances, dynamics, and
  implications.
\newblock In {\em A networked self}. Routledge, 2010, pp.~47--66.

\bibitem{boyd2014s}
{\sc Boyd, D.}
\newblock {\em It's complicated: The social lives of networked teens}.
\newblock Yale University Press, 2014.

\bibitem{boyd2013connected}
{\sc Boyd, D., and Hargittai, E.}
\newblock Connected and concerned: Variation in parents' online safety
  concerns.
\newblock {\em Policy \& Internet 5}, 3 (2013), 245--269.

\bibitem{boyd2011social}
{\sc Boyd, D., and Marwick, A.}
\newblock Social steganography: Privacy in networked publics.
\newblock {\em International Communication Association, Boston, MA 93\/}
  (2011).

\bibitem{cho2016networked}
{\sc Cho, H., and Filippova, A.}
\newblock Networked privacy management in facebook: A mixed-methods and
  multinational study.
\newblock In {\em Conference on Computer-Supported Cooperative Work and Social
  Computing (CSCW)\/} (2016).

\bibitem{cho2018collective}
{\sc Cho, H., Knijnenburg, B., Kobsa, A., and Li, Y.}
\newblock Collective privacy management in social media: A cross-cultural
  validation.
\newblock {\em Transactions on Computer-Human Interaction (TOCHI) 25}, 3
  (2018).

\bibitem{choi2018instagram}
{\sc Choi, T.~R., and Sung, Y.}
\newblock Instagram versus snapchat: Self-expression and privacy concern on
  social media.
\newblock {\em Telematics and Informatics 35}, 8 (2018), 2289--2298.

\bibitem{chou2021teens}
{\sc Chou, H.-L., and Chou, C.}
\newblock How teens negotiate privacy on social media proactively and
  reactively.
\newblock {\em New Media \& Society\/} (2021).

\bibitem{chua2016follow}
{\sc Chua, T. H.~H., and Chang, L.}
\newblock Follow me and like my beautiful selfies: Singapore teenage girls’
  engagement in self-presentation and peer comparison on social media.
\newblock {\em Computers in Human Behavior 55\/} (2016), 190--197.

\bibitem{de2020contextualizing}
{\sc De~Wolf, R.}
\newblock Contextualizing how teens manage personal and interpersonal privacy
  on social media.
\newblock {\em New Media \& Society 22}, 6 (2020), 1058--1075.

\bibitem{dennen2017context}
{\sc Dennen, V.~P., Rutledge, S.~A., Bagdy, L.~M., Rowlett, J.~T., Burnick, S.,
  and Joyce, S.}
\newblock Context collapse and student social media networks: Where life and
  high school collide.
\newblock In {\em Conference on Social media \& Society (SMSOCIETY)\/} (2017).

\bibitem{devito2017platforms}
{\sc DeVito, M.~A., Birnholtz, J., and Hancock, J.~T.}
\newblock Platforms, people, and perception: Using affordances to understand
  self-presentation on social media.
\newblock In {\em Conference on Computer Supported Cooperative Work and Social
  Computing (CSCW)\/} (2017).

\bibitem{devito2018too}
{\sc DeVito, M.~A., Walker, A.~M., and Birnholtz, J.}
\newblock 'too gay for facebook' presenting lgbtq+ identity throughout the
  personal social media ecosystem.
\newblock {\em Proceedings of the ACM on Human-Computer Interaction 2}, CSCW
  (2018), 1--23.

\bibitem{dewar2019finsta}
{\sc Dewar, S., Islam, S., Resor, E., and Salehi, N.}
\newblock Finsta: Creating "fake" spaces for authentic performance.
\newblock In {\em Extended Abstracts of the Conference on Human Factors in
  Computing Systems (CHI EA)\/} (2019), pp.~1--6.

\bibitem{dhir2016adolescents}
{\sc Dhir, A., Kaur, P., Lonka, K., and Nieminen, M.}
\newblock Why do adolescents untag photos on facebook?
\newblock {\em Computers in Human Behavior 55\/} (2016), 1106--1115.

\bibitem{duffy2019you}
{\sc Duffy, B.~E., and Chan, N.~K.}
\newblock "you never really know who’s looking": Imagined surveillance across
  social media platforms.
\newblock {\em New Media \& Society 21}, 1 (2019), 119--138.

\bibitem{emami2018influence}
{\sc Emami~Naeini, P., Degeling, M., Bauer, L., Chow, R., Cranor, L.~F.,
  Haghighat, M.~R., and Patterson, H.}
\newblock The influence of friends and experts on privacy decision making in
  iot scenarios.
\newblock {\em Proceedings of the ACM on Human-Computer Interaction 2}, CSCW
  (2018), 1--26.

\bibitem{tiktok}
{\sc Evans, A., and Sharma, A.}
\newblock Furthering our safety and privacy commitments for teens on tiktok.
\newblock
  \url{https://newsroom.tiktok.com/en-ie/furthering-our-safety-and-privacy-commitments-for-teens-on-tiktok-ire},
  2021.

\bibitem{feng2014teens}
{\sc Feng, Y., and Xie, W.}
\newblock Teens’ concern for privacy when using social networking sites: An
  analysis of socialization agents and relationships with privacy-protecting
  behaviors.
\newblock {\em Computers in Human Behavior 33\/} (2014), 153--162.

\bibitem{fogel2009internet}
{\sc Fogel, J., and Nehmad, E.}
\newblock Internet social network communities: Risk taking, trust, and privacy
  concerns.
\newblock {\em Computers in Human Behavior 25}, 1 (2009), 153--160.

\bibitem{guest2020simple}
{\sc Guest, G., Namey, E., and Chen, M.}
\newblock A simple method to assess and report thematic saturation in
  qualitative research.
\newblock {\em PloS one 15}, 5 (2020), e0232076.

\bibitem{hargittai2010facebook}
{\sc Hargittai, E., et~al.}
\newblock Facebook privacy settings: Who cares?
\newblock {\em First Monday\/} (2010).

\bibitem{hargittai2016can}
{\sc Hargittai, E., and Marwick, A.}
\newblock "what can i really do?" explaining the privacy paradox with online
  apathy.
\newblock {\em International journal of communication 10\/} (2016), 21.

\bibitem{hofstra2016understanding}
{\sc Hofstra, B., Corten, R., and van Tubergen, F.}
\newblock Understanding the privacy behavior of adolescents on facebook: The
  role of peers, popularity and trust.
\newblock {\em Computers in Human Behavior 60\/} (2016), 611--621.

\bibitem{john2013sharing}
{\sc John, N.~A.}
\newblock Sharing and web 2.0: The emergence of a keyword.
\newblock {\em New media \& society 15}, 2 (2013), 167--182.

\bibitem{kang2021privacy}
{\sc Kang, C.}
\newblock Lawmakers urge the head of instagram to better protect children.
\newblock
  \url{https://www.nytimes.com/2021/12/08/technology/adam-mosseri-instagram-senate.html},
  2021.

\bibitem{kang2021teens}
{\sc Kang, H., Shin, W., and Huang, J.}
\newblock Teens' privacy management on video-sharing social media: the roles of
  perceived privacy risk and parental mediation.
\newblock {\em Internet Research\/} (2021).

\bibitem{lang2015just}
{\sc Lang, C., and Barton, H.}
\newblock Just untag it: Exploring the management of undesirable facebook
  photos.
\newblock {\em Computers in Human Behavior 43\/} (2015), 147--155.

\bibitem{pew2015teens}
{\sc Lenhart, A.}
\newblock Teens, social media and technology overview 2015.
\newblock Tech. rep., Pew Research Center, 2015.

\bibitem{livingstone2008taking}
{\sc Livingstone, S.}
\newblock Taking risky opportunities in youthful content creation: teenagers'
  use of social networking sites for intimacy, privacy and self-expression.
\newblock {\em New Media \& Society 10}, 3 (2008), 393--411.

\bibitem{livingstone2014developing}
{\sc Livingstone, S.}
\newblock Developing social media literacy: How children learn to interpret
  risky opportunities on social network sites.
\newblock {\em Communications 39}, 3 (2014), 283--303.

\bibitem{locke2010eavesdropping}
{\sc Locke, J.~L.}
\newblock {\em Eavesdropping: An intimate history}.
\newblock OUP Oxford, 2010.

\bibitem{ma2016anonymity}
{\sc Ma, X., Hancock, J., and Naaman, M.}
\newblock Anonymity, intimacy and self-disclosure in social media.
\newblock In {\em Conference on Human Factors in Computing Systems (CHI)\/}
  (2016).

\bibitem{pew2013teens}
{\sc Madden, M., Lenhart, A., Cortesi, S., Gasser, U., Duggan, M., Smith, A.,
  and Beaton, M.}
\newblock Teens, social media, and privacy.
\newblock {\em Pew Research Center 21}, 1055 (2013), 2--86.

\bibitem{mansour2021collective}
{\sc Mansour, A., and Francke, H.}
\newblock Collective privacy management practices: A study of privacy
  strategies and risks in a private facebook group.
\newblock {\em Proceedings of the ACM on Human-Computer Interaction 5}, CSCW
  (2021), 1--27.

\bibitem{marwick2011tweet}
{\sc Marwick, A.~E., and Boyd, D.}
\newblock I tweet honestly, i tweet passionately: Twitter users, context
  collapse, and the imagined audience.
\newblock {\em New Media \& Society 13}, 1 (2011), 114--133.

\bibitem{marwick2014networked}
{\sc Marwick, A.~E., and boyd, d.}
\newblock Networked privacy: How teenagers negotiate context in social media.
\newblock {\em New Media \& Society 16}, 7 (2014), 1051--1067.

\bibitem{may2014safeguarding}
{\sc May-Chahal, C., Mason, C., Rashid, A., Walkerdine, J., Rayson, P., and
  Greenwood, P.}
\newblock Safeguarding cyborg childhoods: Incorporating the on/offline
  behaviour of children into everyday social work practices.
\newblock {\em British Journal of Social Work 44}, 3 (2014), 596--614.

\bibitem{mchugh2018social}
{\sc McHugh, B.~C., Wisniewski, P., Rosson, M.~B., and Carroll, J.~M.}
\newblock When social media traumatizes teens: The roles of online risk
  exposure, coping, and post-traumatic stress.
\newblock {\em Internet Research\/} (2018).

\bibitem{mendel2017susceptibility}
{\sc Mendel, T., and Toch, E.}
\newblock Susceptibility to social influence of privacy behaviors: Peer versus
  authoritative sources.
\newblock In {\em Proceedings of the 2017 ACM Conference on Computer Supported
  Cooperative Work and Social Computing\/} (2017), pp.~581--593.

\bibitem{oolo2013performing}
{\sc Oolo, E., and Siibak, A.}
\newblock Performing for one’s imagined audience: Social steganography and
  other privacy strategies of estonian teens on networked publics.
\newblock {\em Cyberpsychology: Journal of Psychosocial Research on Cyberspace
  7}, 1 (2013).

\bibitem{palen2003unpacking}
{\sc Palen, L., and Dourish, P.}
\newblock Unpacking" privacy" for a networked world.
\newblock In {\em Conference on Human Factors in Computing Systems (CHI)\/}
  (2003).

\bibitem{papacharissi2011fifteen}
{\sc Papacharissi, Z., and Gibson, P.~L.}
\newblock Fifteen minutes of privacy: Privacy, sociality, and publicity on
  social network sites.
\newblock In {\em Privacy Online}. Springer, 2011, pp.~75--89.

\bibitem{petronio2002boundaries}
{\sc Petronio, S.}
\newblock {\em Boundaries of privacy: Dialectics of disclosure}.
\newblock Suny Press, 2002.

\bibitem{rashidi2020s}
{\sc Rashidi, Y., Kapadia, A., Nippert-Eng, C., and Su, N.~M.}
\newblock "it's easier than causing confrontation": Sanctioning strategies to
  maintain social norms and privacy on social media.
\newblock {\em Proceedings of the ACM on Human-Computer Interaction 4}, CSCW1
  (2020), 1--25.

\bibitem{salter2016privates}
{\sc Salter, M.}
\newblock Privates in the online public: Sex (ting) and reputation on social
  media.
\newblock {\em New Media \& Society 18}, 11 (2016), 2723--2739.

\bibitem{santana2014virtuous}
{\sc Santana, A.~D.}
\newblock Virtuous or vitriolic: The effect of anonymity on civility in online
  newspaper reader comment boards.
\newblock {\em Journalism practice 8}, 1 (2014), 18--33.

\bibitem{sciara2021going}
{\sc Sciara, S., Contu, F., Bianchini, M., Chiocchi, M., and Sonnewald, G.~G.}
\newblock Going public on social media: The effects of thousands of instagram
  followers on users with a high need for social approval.
\newblock {\em Current Psychology\/} (2021), 1--15.

\bibitem{scolere2018constructing}
{\sc Scolere, L., Pruchniewska, U., and Duffy, B.~E.}
\newblock Constructing the platform-specific self-brand: The labor of social
  media promotion.
\newblock {\em Social Media + Society 4}, 3 (2018).

\bibitem{strano2013covering}
{\sc Strano, M.~M., and Queen, J.~W.}
\newblock Covering your face on facebook.
\newblock {\em Journal of Media Psychology\/} (2013).

\bibitem{stutzman2010friends}
{\sc Stutzman, F., and Kramer-Duffield, J.}
\newblock Friends only: examining a privacy-enhancing behavior in facebook.
\newblock In {\em Conference on Human Factors in Computing Systems (CHI)\/}
  (2010).

\bibitem{sunden2003material}
{\sc Sund{\'e}n, J.}
\newblock {\em Material virtualities: Approaching online textual embodiment}.
\newblock Peter Lang, 2003.

\bibitem{taber2018personality}
{\sc Taber, L., and Whittaker, S.}
\newblock Personality depends on the medium: differences in self-perception on
  snapchat, facebook and offline.
\newblock In {\em Conference on Human Factors in Computing Systems (CHI)\/}
  (2018).

\bibitem{taber2020finsta}
{\sc Taber, L., and Whittaker, S.}
\newblock " on finsta, i can say'hail satan'": Being authentic but disagreeable
  on instagram.
\newblock In {\em Conference on Human Factors in Computing Systems (CHI)\/}
  (2020).

\bibitem{trieu2020private}
{\sc Trieu, P., and Baym, N.~K.}
\newblock Private responses for public sharing: understanding self-presentation
  and relational maintenance via stories in social media.
\newblock In {\em Conference on Human Factors in Computing Systems (CHI)\/}
  (2020).

\bibitem{tufekci2008can}
{\sc Tufekci, Z.}
\newblock Can you see me now? audience and disclosure regulation in online
  social network sites.
\newblock {\em Bulletin of Science, Technology \& Society 28}, 1 (2008),
  20--36.

\bibitem{tufekci2012facebook}
{\sc Tufekci, Z.}
\newblock Facebook, youth and privacy in networked publics.
\newblock In {\em The International Conference on Weblogs and Social Media
  (ICWSM)\/} (2012).

\bibitem{wells2021fb}
{\sc Wells, G., Horwitz, J., and Seetharaman, D.}
\newblock Facebook knows instagram is toxic for teen girls, company documents
  show.
\newblock
  \url{https://www.wsj.com/articles/facebook-knows-instagram-is-toxic-for-teen-girls-company-documents-show-11631620739},
  2021.

\bibitem{wisniewski2016framing}
{\sc Wisniewski, P., Islam, A., Richter~Lipford, H., and Wilson, D.~C.}
\newblock Framing and measuring multi-dimensional interpersonal privacy
  preferences of social networking site users.
\newblock {\em Communications of the Association for information systems 38}, 1
  (2016), 10.

\bibitem{xiao2020random}
{\sc Xiao, S., Metaxa, D., Park, J.~S., Karahalios, K., and Salehi, N.}
\newblock Random, messy, funny, raw: finstas as intimate reconfigurations of
  social media.
\newblock In {\em Conference on Human Factors in Computing Systems (CHI)\/}
  (2020).

\bibitem{xie2015see}
{\sc Xie, W., and Kang, C.}
\newblock See you, see me: Teenagers’ self-disclosure and regret of posting
  on social network site.
\newblock {\em Computers in Human Behavior 52\/} (2015), 398--407.

\bibitem{yang2016exploring}
{\sc Yang, K.~C., Pulido, A., and Yowei, K.}
\newblock Exploring the relationship between privacy concerns and social media
  use among college students: A communication privacy management perspective.
\newblock {\em Intercultural Communication Studies 25}, 2 (2016).

\bibitem{yau2019s}
{\sc Yau, J.~C., and Reich, S.~M.}
\newblock “it's just a lot of work”: Adolescents’ self-presentation norms
  and practices on facebook and instagram.
\newblock {\em Journal of research on adolescence 29}, 1 (2019), 196--209.

\bibitem{zhang2016nosy}
{\sc Zhang-Kennedy, L., Mekhail, C., Abdelaziz, Y., and Chiasson, S.}
\newblock From nosy little brothers to stranger-danger: Children and parents'
  perception of mobile threats.
\newblock In {\em ACM International Conference on Interaction Design and
  Children\/} (2016).

\bibitem{zhao2016social}
{\sc Zhao, X., Lampe, C., and Ellison, N.~B.}
\newblock The social media ecology: User perceptions, strategies and
  challenges.
\newblock In {\em Conference on Human Factors in Computing Systems (CHI)\/}
  (2016).

\end{thebibliography}
\begin{flushleft}
\textbf{{\huge Appendix}}
\end{flushleft}
\appendix
\section{Interview Protocol}
\label{sec:interview}
\textbf{Introduction Questions}

Hi, thanks for coming to talk to us today. The theme of our research is really to see how young adults think about privacy on social media. We’re excited to hear more about how you navigate think about privacy on Instagram and different features you may use. I might ask follow-up questions for clarification as well. Please answer honestly to every question, there are no wrong answers.

[For under 18 participants] We have received both your parental permission and assent form for the interview. 

[For over 18 participants] We have received your consent form. 

[If they agreed to audiorecording] Based on the assent / consent form, you have agreed to have us audiorecord this interview. Again, these recordings will only be used for research purposes. If you would rather not be recorded, please let us know now. 

If you want to stop at any time or take a break, we can. If we ask a question you don’t want to answer, please feel free to let us know. That is absolutely okay.

As stated in the consent / assent form, during the study, we may ask you to share an example of an Instagram post if you feel comfortable. If you are not comfortable, you can opt to not share this with us.

\begin{enumerate}
    \item Can you tell us a little bit more about yourself? How old are you / what grade are you in? What do you do for fun? 
\end{enumerate}

\textbf{General Instagram Overview}
\begin{enumerate}
  \item Can you tell us a little bit about how you use Instagram? What do you mainly use it for? 
  \item If you feel comfortable, do you mind showing us what your Instagram account looks like? 
  \item Do you have any information on Instagram that is available for anyone to see? 
\end{enumerate}

\textbf{Understanding General Privacy Usage}
\begin{enumerate}
  \item What kind of things do you post on your Instagram? Looking at your most recent Instagram post, walk me through your thought process when posting this.
  \item	Who do you want these posts to be seen by? 	Do you ever post content for a specific audience? 
  \item Are there people who you would rather not see your posts?
  \item How do you prevent people from seeing your posts?
  \begin{enumerate}
    \item Have you ever had an instance when someone saw something you posted that you didn’t want them to? How often does this occur? What did you do in response? 
  \end{enumerate}
  \item How do you feel about being in other people’s posts? Do you want to be tagged in them?
  \item How do you feel about people commenting on your posts? 
    \begin{enumerate}
    \item What type of content do people usually comment on your posts? Who usually comments? 
    \item What about when you comment on other people’s posts? Whose posts do you comment on? 
    \end{enumerate}
  \item Have you used any of Instagram’s features to make things private?
\end{enumerate}

\textbf{Questions for Specific Features}
\begin{enumerate}
    \item What other features do you use on Instagram? 
    \item Have you ever made a story on Instagram?         \begin{enumerate}
            \item Can you walk me through the process on how you decided to post your most recent story?
            \item How do you choose what to story? 
            \item Why did you choose to story it versus post? 
        \end{enumerate}
    \item Have you ever pinned a collection of stories on Instagram?
        \begin{enumerate}
            \item If no, why not? 
            \item If so, what kinds of collections do you make? 
            \item How do you decide what stories should be put in collections? 
        \end{enumerate}
    \item Have you ever used Close Friends on Instagram? 
        \begin{enumerate}
            \item How did you choose your Close Friends? 
            \item Can you give me an example of content you shared with only Close Friends?
        \end{enumerate}
    \item How many Instagram accounts do you have? 
        \begin{enumerate}
            \item How do you choose what to post on each account?
            \item Do you have a main account and why did you choose to create the additional account(s)? 
            \item Can you give me an example of content you posted on one of your additional accounts and why you chose to post it on that account?
        \end{enumerate}    
    \item How do you use DMs? 
        \begin{enumerate}
            \item Who do you usually DM? 
            \item What kind of things do you DM? 
            \item Why do you choose to DM rather than commenting?
        \end{enumerate}
\end{enumerate}

\textbf{Effectiveness of Privacy Features}
\begin{enumerate}
    \item Could you tell me about a time when you removed a post? What was it? 
    \begin{enumerate}
        \item Why did you remove it?
        \item If you have never removed a post, do you know anyone who has? 
    \end{enumerate}
    \item Could you tell me about a time when you used editing tools to remove certain content from an image? 
    \item Could you tell me about a time you were tagged in a post you didn’t want to be in? 
        \begin{enumerate}
        \item What did you do in response? Why? 
        \item Can you give me an example of a post you wouldn’t want to be tagged in?
    \end{enumerate}
    \item Have you ever changed the privacy status (i.e., public to private) of your Instagram? 
    \item Have you ever deleted and then reopened your Instagram account?  
\end{enumerate}

\textbf{Understanding in relation to other social media platforms}
\begin{enumerate}
    \item Do you have other social media accounts? How do you choose what to share on Instagram versus other social media?
\end{enumerate}

\textbf{Conclusion Questions}
\begin{enumerate}
    \item Is there anything I should have asked you, or anything else you want to share? 
\end{enumerate}

\section{Survey Questions} 
\label{sec:survey}
\begin{enumerate}
    \item How many Instagram accounts do you have?
    \item Is your main Instagram account private or public? 
    \item How long have you had an Instagram account? If you have multiple, the longest you have had any account.
    \item How many posts? Followers? Following?
    \item Of the last 10 people you have followed, how many have you interacted with in person?
    \item How often have you used Instagram Explore in the past week? 
    \item On average, how long are you on Instagram per day?
    \item Which of the following features have you used on Instagram?
    \item Are there other features you use on Instagram we have not listed?
    \item Have you used any of the following (Close Friends on Instagram, Snapchat private stories, Finsta) currently? 
    \item Have you used any of the following (Close Friends on Instagram, Snapchat private stories, Finsta) in the past (but not currently)? 
    \item Who do you allow to follow or view (e.g., parents, mutuals, acquaintances, friends, romantic partner, best friends) content on your main Instagram, finsta, Close Friends story, and Snapchat private story? 
    \item Why have you untagged yourself in a post? 
    \item Why have your archived or deleted a post? 
    \item Of your 10 most recent posts, how many were previously archived posts that you then unarchived? 
    \item When have you asked for someone else's permission before sharing a post / story? 
    \item What do you include in your Instagram bio? 
    \item Of your 10 most recent posts, how many were previously archived posts that you then unarchived? 
    \item Of your 10 last stories, how many were related to political or social issues?
    \item Did you find out about the following features (e.g., Close Friends, untagging, archiving, blocking followers, restricting followers, bookmarking posts, blocking someone from your story, removing a comment) via someone else's post / story, word of mouth, learned by yourself, or did you not know about this feature?  
    \item How old are you?
    \item What is your gender?
    \item What is your race / ethnicity?
    \item Would you be interested in participating in an interview?
\end{enumerate}

\section{Survey Results}
\label{sec:survey_results}
\subsection{Participant demographics}
\begin{figure}
    \centering
    \includegraphics[width=0.8\textwidth]{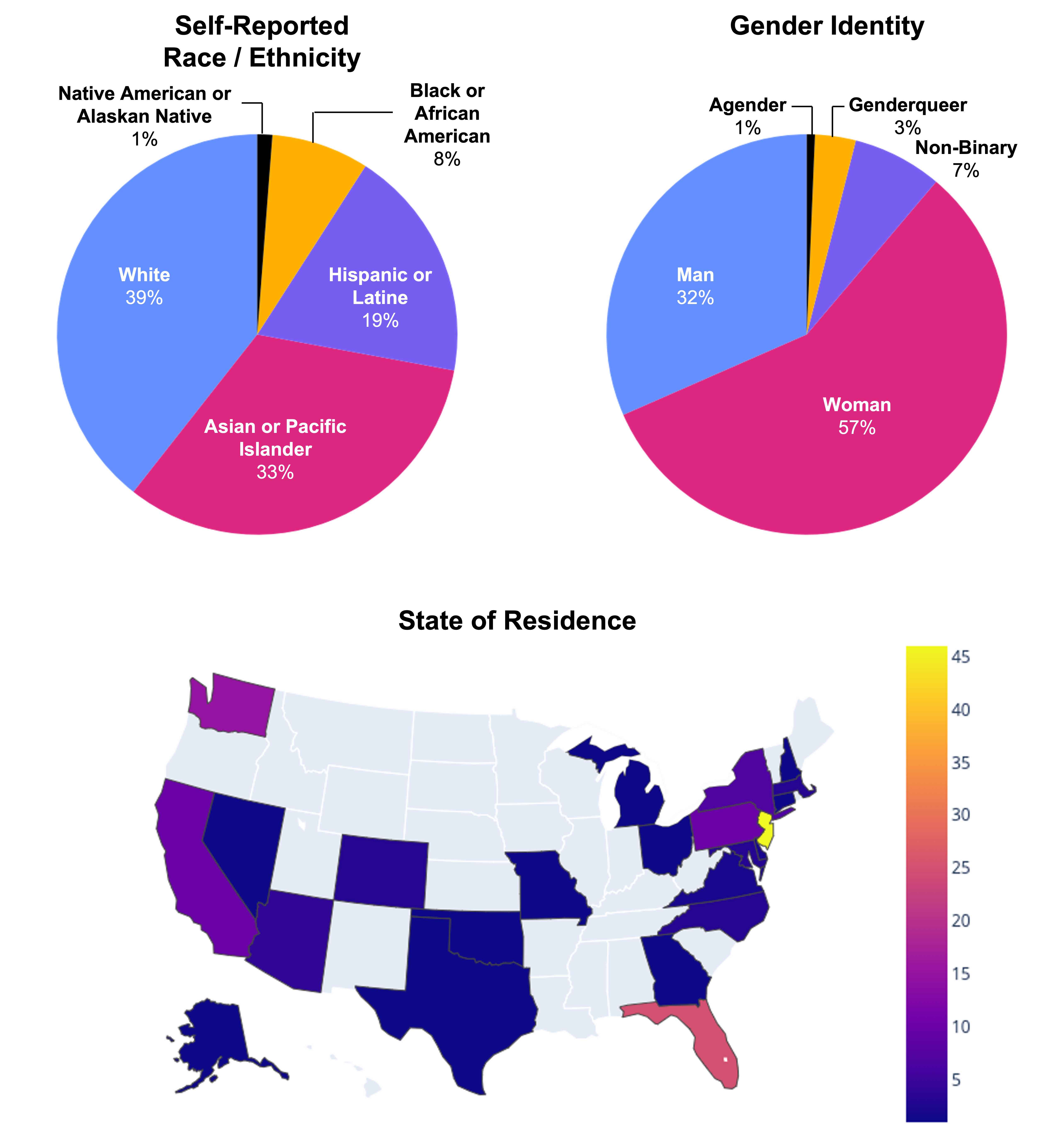}
    \caption{Demographic information (e.g., race, age, and state) for all survey respondents}
    \label{fig:surv_dem}
\end{figure}

First, we provide more details on the participant demographics (N=144) results from our online survey (Sec. 3.1). As shown in Fig.~\ref{fig:surv_dem}, the majority ($57\%$) of our respondents identified as women. $39\%$ of respondents identified as White, $33\%$ as Asian or Pacific Islander, $19\%$ as Hispanic or Latine, $8\%$ as Black or African American, and $1\%$ as Native American or Alaskan Native. In terms of location, we had respondents from 23 states with the most popular locations being New Jersey (N=46), Florida (N=25), Pennsylvania (N=10), and Washington (N=10). The average age of survey respondents was 16.8 years old.

We also look into information about account usage on Instagram. We found that the majority (N = 83) of participants had more than one Instagram account. On average, participants have had their accounts for $3.64$ years with 59 respondents having had their main account for five or more years. 

\subsection{Channels for learning about privacy features}
In Sec.~\ref{sec:peer}, we discuss how participants learn about privacy features after observing their peers. In our survey, we ask participants how they learned about different privacy features on Instagram (e.g., learned by themselves, from observing a peer's account, word of mouth, or did not know about feature). Specifically, we ask about Close Friends stories, untagging, archiving, blocking followers, restricting followers, bookmarking posts, blocking someone from your story, and removing a comment.

\begin{figure}[t]
    \centering
    \includegraphics[width=0.95\textwidth]{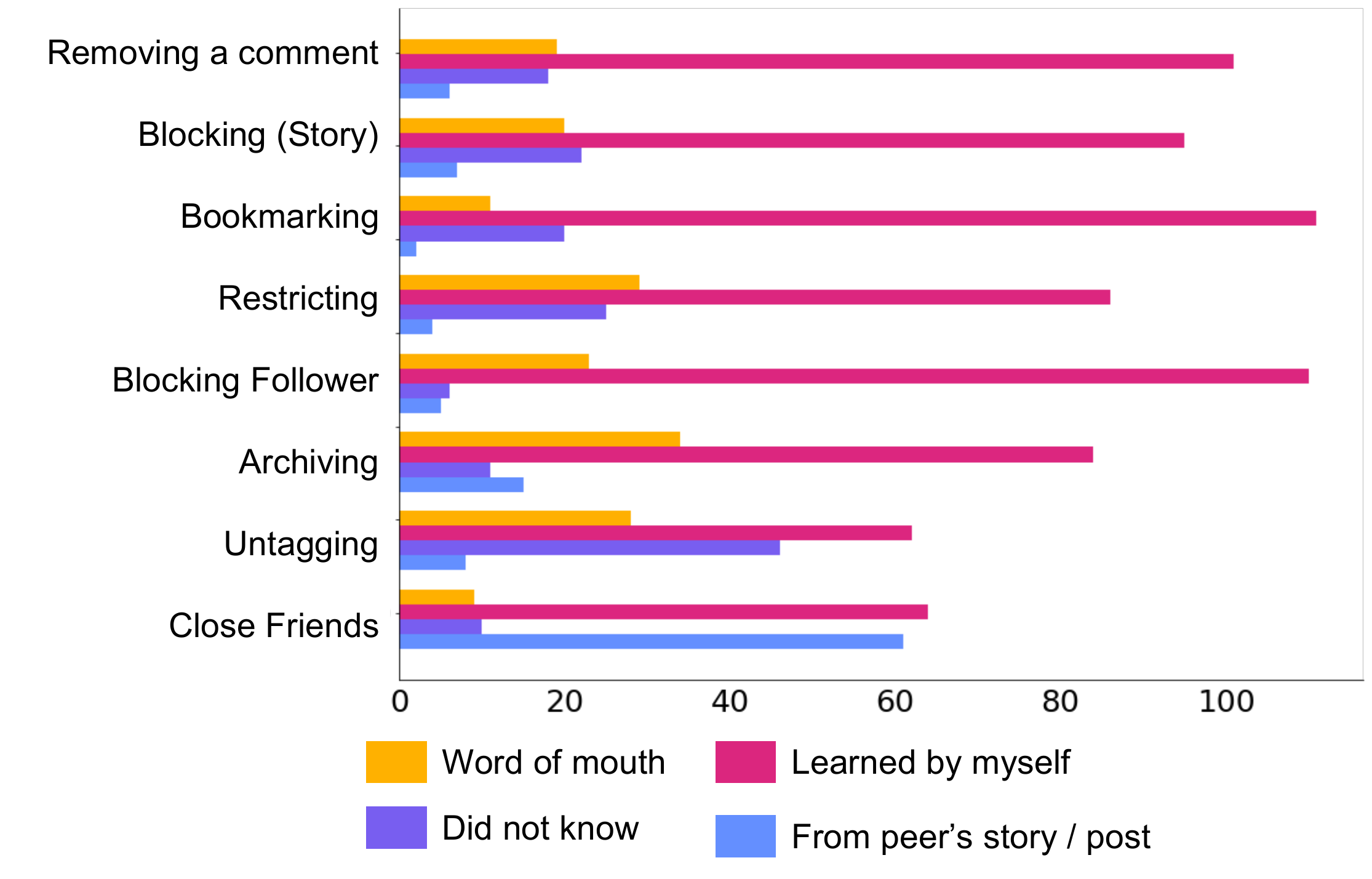}
    \caption{Visualization of how survey respondents learned about various privacy features on Instagram}
    \label{fig:learning}
\end{figure}

From Fig.~\ref{fig:learning}, we make three key observations.  First, as expected, we find that across the features, the majority of respondents are learning about features on their own. Second, we do find that peer-to-peer pathways are a fruitful way of learning about different features. Across the eight privacy features, on average, respondents learned about the features through word of mouth or other people's stories / posts $24.9\%$ of the time. Certain features, such as Close Friends ($48.6\%$) and archiving ($34.0\%$), were more likely to spend via peer-to-peer pathways, than others, such as bookmarking posts ($9.0\%$). Finally, as observed in Sec. 5.3.4, the respondents were unaware of many of the privacy features available to them. Across the features, on average $13.7\%$ of respondents did not know about the feature. In fact, $32.0\%$ of respondents did not know about untagging as a feature.

\subsection{Reasons for untagging and archiving posts} 
In Sec.~\ref{sec:body}, we discuss reasons that respondents archived or deleting posts they made and untagging themselves from posts. We find that, in both cases, not liking the way they looked was the most comment reason for our teen respondents. For archiving, other common rationale including looking too childish in the post and having the post not match the feed aesthetic. Further, we find that while archived images can be unarchived, most respondents view this as a permanent process. Of respondents' most recent 10 posts, on average only 0.91 was previously archived content. Looking at untagging, we find that respondents were less likely to use this feature. Other common reasons for untagging included not knowing the tagger and no longer being friends with the person who tagged them in the post.
\begin{figure}
    \centering
    \includegraphics[width=0.95\textwidth]{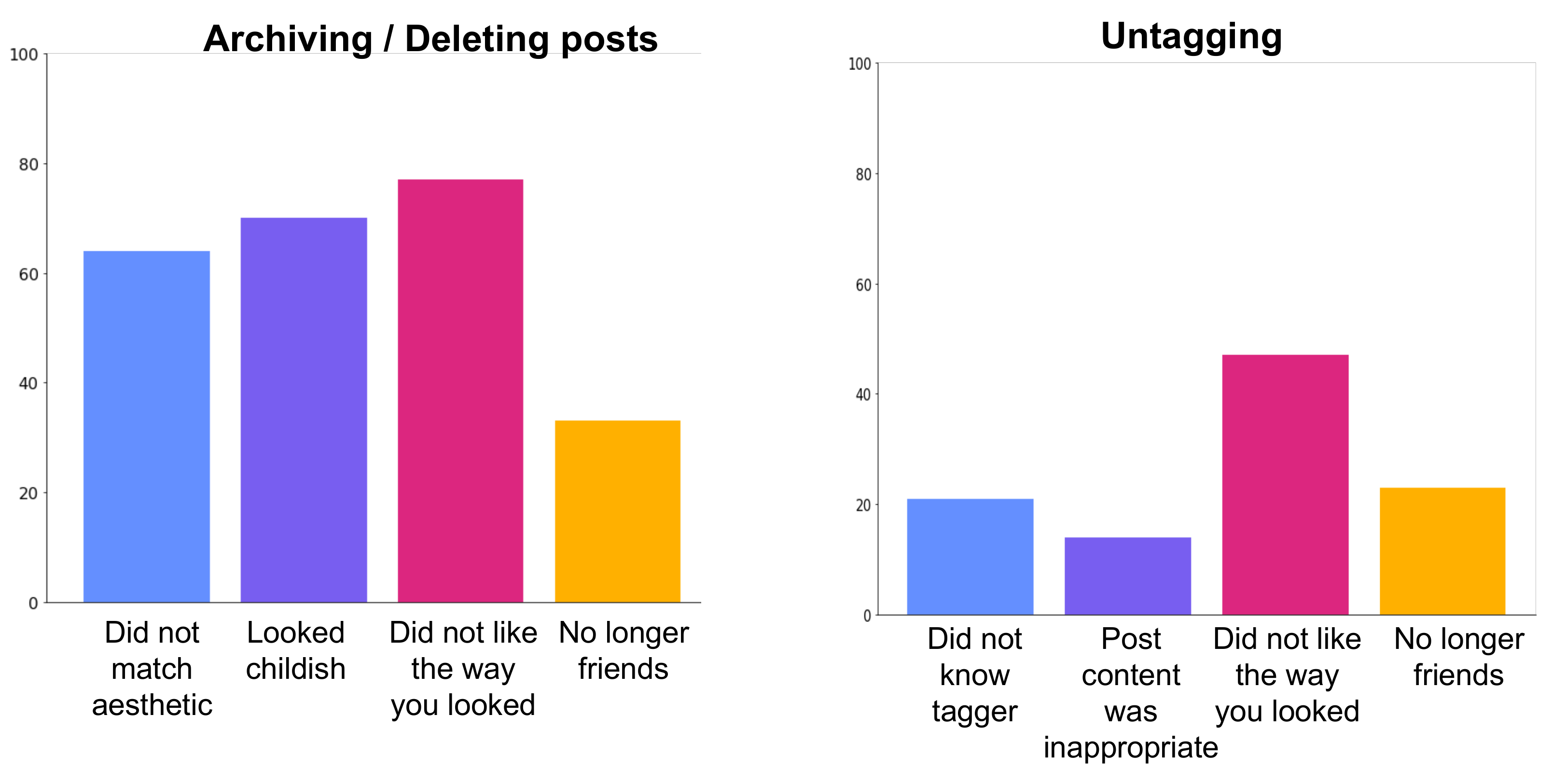}
    \caption{Visualization of why survey respondents archived or deleted posts (left) or untagged themselves from posts (right)}
    \label{fig:untagging}
\end{figure}

\end{document}